\documentclass[useAMS,usenatbib,usegraphicx]{mn2e}

\usepackage[latin1]{inputenc}
\usepackage{amsmath}
\usepackage{amsfonts}
\usepackage{amssymb}
\usepackage{color}

\usepackage{times}
\usepackage[colorlinks=false,dvips]{hyperref}

\newcommand{\Msun}{\ensuremath{\textrm{M}_{\odot}}}
\newcommand{\Rsun}{\ensuremath{\textrm{R}_{\odot}}}
\newcommand{\Lsun}{\ensuremath{\textrm{L}_{\odot}}}
\newcommand{\lbol}{\ensuremath{L_\textrm{bol}}}
\newcommand{\kms}{km\hspace{0.25em}s$^{-1}$}

\newcommand{\HI}{\mbox{H\hspace{0.25em}{\sc i}}}

\newcommand{\OI}{\mbox{O\hspace{0.25em}{\sc i}}}

\newcommand{\CII}{\mbox{C\hspace{0.25em}{\sc ii}}}

\newcommand{\SII}{\mbox{S\hspace{0.25em}{\sc ii}}}

\newcommand{\SiII}{\mbox{Si\hspace{0.25em}{\sc ii}}}
\newcommand{\SiIII}{\mbox{Si\hspace{0.25em}{\sc iii}}}
\newcommand{\CaII}{\mbox{Ca\hspace{0.25em}{\sc ii}}}

\newcommand{\Fefs}{$^{56}$Fe}
\newcommand{\Feffo}{$^{54}$Fe}
\newcommand{\Cofs}{$^{56}$Co}
\newcommand{\Nifs}{$^{56}$Ni}

\newcommand{\KE}{\ensuremath{E_K}}
\newcommand{\Dm}{\ensuremath{\Delta m_{15}(B)}}

\newcommand{\aap}{A\&A}

\newcommand{\apj}{ApJ}
\newcommand{\apjs}{ApJS}
\newcommand{\apjl}{ApJ}
\newcommand{\aj}{AJ}

\newcommand{\mnras}{MNRAS}
\newcommand{\na}{New Astronomy}
\newcommand{\nat}{Nature}

\newcommand{\pasp}{PASP}

\newcommand{\sci}{Sci}

\newcommand{\araa}{ARA\&A}

\newcommand{\col}{\ensuremath{\mathcal{C}}}
\newcommand{\tr}{\ensuremath{t_\textrm{r}}}
\newcommand{\eg}{e.g.\ }
\defcitealias{maguire12}{M12}
\defcitealias{nugent11}{N11}
\defcitealias{bloom12}{B12}

\definecolor{myorange}{rgb}{0.9,0.6,0.0}

\definecolor{myredder}{rgb}{1.0,0.0,0.0}

\definecolor{mygreen}{rgb}{0.0,0.7,0.0}

\definecolor{myblue}{rgb}{0.0,0.0,1.0}

\definecolor{myyellow}{rgb}{0.92,0.89,0.00}

\begin{document}

\title[SN\,2011fe: models for \textit{HST} UV spectra]
  {\textit{Hubble Space Telescope} spectra of the type Ia supernova SN\,2011fe:
  a tail of low-density, high-velocity material with $Z<Z_{\odot}$.}
\author[P. A.~Mazzali et al.]{P. A. Mazzali$^{1,2,3}$
\thanks{E-mail: P.Mazzali@ljmu.ac.uk}, 
M. Sullivan$^{4}$, S. Hachinger$^{2,5}$, R. S. Ellis$^{6}$, P. E. Nugent$^{7,8}$, 
\newauthor D. A. Howell$^{9,10}$, A. Gal-Yam$^{11}$, K. Maguire$^{12}$, 
J. Cooke$^{13}$, R. Thomas$^{8}$, K. Nomoto$^{14}$,
\newauthor  E. S. Walker$^{15}$\\
\\
  $^{1}$Astrophysics Research Institute, Liverpool John Moores University, IC2, Liverpool Science Park, 146 Brownlow Hill, Liverpool L3 5RF, UK\\
  $^{2}$Istituto Nazionale di Astrofisica-OAPd, vicolo dell'Osservatorio 5, 35122 Padova, Italy\\
  $^{3}$Max-Planck-Institut für Astrophysik, Karl-Schwarzschild-Str. 1, D-85748 Garching, Germany\\
  $^{4}$School of Physics and Astronomy, University of Southampton, Southampton, SO17 1BJ, UK\\
  $^{5}$Institut f\"ur Theoretische Physik und Astrophysik, Universit\"at W\"urzburg, Emil-Fischer-Str. 31, 97074 W\"urzburg, Germany\\
  $^{6}$Cahill Center for Astrophysics, California Institute of Technology, Pasadena, CA 91125, USA\\
  $^{7}$Department of Astronomy, University of California, Berkeley, CA 94720-3411, USA\\
  $^{8}$Computational Cosmology Center, Lawrence Berkeley National Laboratory, 1 Cyclotron Road, Berkeley, CA 94720, USA\\
  $^{9}$Las Cumbres Observatory Global Telescope Network, Goleta, CA 93117, USA\\
  $^{10}$Department of Physics, University of California, Santa Barbara, CA 93106-9530, USA\\
  $^{11}$Benoziyo Center for Astrophysics, Weizmann Institute of Science, 76100 Rehovot, Israel\\
  $^{12}$Department of Physics (Astrophysics), University of Oxford, Keble Road, Oxford OX1 3RH\\
  $^{13}$Centre for Astrophysics \& Supercomputing, Swinburne University of Technology, Mail H30, PO Box 218, Hawthorn, Victoria 3122, Australia\\
  $^{14}$Kavli IPMU, University of Tokyo, Kashiwanoha 5-1-5, Kashiwa, Chiba 277-8583, Japan\\
  $^{15}$Yale University, Department of Physics, P.O. box 208120, New Haven, CT 06520-8120, USA \\
}

\date{Accepted ... Received ...; in original form ...}
\pubyear{2013}
\volume{}
\pagerange{}

\maketitle

\begin{abstract}
  \textit{Hubble Space Telescope} spectroscopic observations of the nearby type
  Ia supernova (SN Ia) SN\,2011fe, taken on 10 epochs from $-$13.1 to $+$40.8
  days relative to $B$-band maximum light, and spanning the far-ultraviolet (UV)
  to the near-infrared (IR) are presented.  This spectroscopic coverage makes 
  SN\,2011fe the best-studied local SN\,Ia to date.  SN\,2011fe is a typical
  moderately-luminous SN Ia with no evidence for dust extinction. Its near-UV
  spectral properties are representative of a larger sample of local events
  \citep{maguire12}. The near-UV to optical spectra of SN\,2011fe are modelled
  with a Monte Carlo radiative transfer code using the technique of `abundance
  tomography', constraining the density structure and the abundance
  stratification in the SN ejecta. SN\,2011fe was a relatively weak explosion,
  with moderate Fe-group yields.  The density structures of the classical model
  W7 and of a delayed  detonation model were tested. Both have shortcomings.  An
  ad-hoc density distribution was developed which yields improved fits and is
  characterised by a high-velocity tail, which is absent in W7.  However, this
  tail contains less mass than delayed detonation models. This improved model
  has a lower energy than one-dimensional explosion models matching typical
  SNe\,Ia \citep[\eg W7, WDD1,][]{iwa99}. The derived Fe abundance in the
  outermost layer is consistent with the metallicity at the SN explosion site
  in M101 ($\sim 0.5 Z_{\odot}$). The spectroscopic rise time ($\sim 19$\,days)
  is significantly longer than that measured from the early optical light curve,
  implying a `dark phase' of $\sim 1$ day. A longer rise time has significant
  implications when deducing the properties of the white dwarf and binary system
  from the early photometric behaviour.
\end{abstract}


\section{Introduction}
\label{sec:introduction}

Type Ia supernovae (SNe Ia) remain a well-exploited cosmological probe and an
immediate route to understanding the nature of dark energy
\citep[e.g.][]{rie98,per99,rie07,kes09,sul11,suz12}. However, the exact nature
of the progenitor system is still not well understood, with various
astrophysical systematics potentially affecting cosmological measurements
\citep[for summaries, see][]{conley11,howell11,astier12}. A particularly
important systematic is the role of the metallicity, or composition, of the
exploding white dwarf star on the outcome of the explosion and the properties of
SNe Ia.

It has long been known that the ultraviolet (UV) region of SN Ia spectra
effectively traces compositional or metallicity effects \citep{hoe98,lentz00a}.
Although the physics underlying the formation of SN Ia UV spectra is complex,
the UV is probably the best diagnostic of iron-group element abundances in the
outer layers of the SN ejecta \citep{maz00,wal12,hachinger13} which are
transparent to optical photons.  This is closely connected to the density
profile of the SN ejecta, and hence to the SN explosion model -- the iron-group
content of incompletely burned zones is one of the main differences between SN
Ia explosions at different metallicity \citep{iwa99}.

However, data in the UV were historically difficult to obtain. Space
observatories are required to perform the observations, and the intrinsic
faintness of SNe Ia below $\sim$2700\AA\ requires targets to be relatively
nearby to obtain a useful signal-to-noise (S/N). The \textit{International
Ultraviolet Explorer} (\textit{IUE}) and the \textit{Hubble Space Telescope}
(\textit{HST}) have both been used
\citep{branch86,jeffrey92,kirshner93,cappellaro95,sauer08}, but generally the
data suffer from poor S/N, with only one well-observed SN Ia near maximum light
\citep[SN\,1992A;][]{kirshner93} and only a few events with early pre-maximum
light coverage \citep[e.g. SN\,1990N;][]{leibundgut91}.  A full compilation of
observations up to 2004 can be found in \citet{foley08b}. More recently, the
Advanced Camera for Surveys on \textit{HST} observed four SNe Ia \citep{wang12},
but the spectral resolution was low making a detailed study difficult.

Some of these earlier difficulties have been alleviated in modern SN searches,
which can provide a significant number of early SN\,Ia events. The Lick
Observatory Supernova Search early discovery of SN\,2009ig
\citep{foley12sn2009ig} allowed a reasonably high-S/N UV spectral time series to
be obtained using \textit{Swift}, and the fast response of that satellite has
also allowed spectral series of several other events, although the S/N below
2700\AA\ is not high \citep{bufano09}. \textit{HST}, with its larger aperture,
has obtained more spectra with a higher S/N, including high-S/N observations of
the relatively fast-declining SN\,2011iv \citep{foley12sn2011iv}, although these
did not commence until maximum light.

A different approach is to focus on the near-UV region down to
$\simeq2900$\,\AA, which is more easily accessible. At high redshift this
wavelength range is redshifted into the optical \citep[e.g.][]{ellis08}, and a
significant sample of single maximum-light near-UV spectra is now available
\citep{ellis08,foley08a,balland09,foley12sdss}. In \textit{HST} cycle 17, we
successfully obtained single-epoch near-UV spectra of $\sim$30 lower redshift
($z<0.08$) SNe Ia at phases close to maximum light \citep[][hereafter
M12]{maguire12}, for the most part using early discoveries from the Palomar
Transient Factory \citep[PTF;][]{rau09,law09}.

\citetalias{maguire12} presented evidence for evolution in the mean near-UV
continuum between $z=0$ and $z\sim0.5$, the latter using spectra from
\citet{ellis08}. They demonstrated an excess near-UV flux at $\sim0.5$ compared
to $z=0$ at $\simeq3$$\sigma$, and showed that this was qualitatively consistent
with expected evolutionary effects due to changes in metallicity via a
comparison to SN Ia models \citep{wal12}. Even stronger near-UV spectral
evolution in the same sense as \citetalias{maguire12} was presented by
\citet{foley12sdss}, but over a narrower range in redshift, comparing a
ground-based $z\simeq0$ spectrum and a mean spectrum at $z\sim0.2$. This would
then imply a puzzlingly inconsistent evolution between $z=0.2$ and $z=0.5$ --
although \citet{foley12sdss} caution that selection effects in the $z\sim0.2$
sample may play a role in this discrepancy.

In cycle 18, we extended the \citetalias{maguire12} programme to obtain
multi-epoch \textit{HST} SN Ia spectra of four events, from the earliest
possible epoch to post-maximum light. Our first spectral series, of SN\,2010jn,
and its detailed analysis using a radiative transfer code are published in
\citet{hachinger13}. SN\,2010jn was an energetic SN Ia explosion, with high
expansion velocities and significant amounts of iron-group elements in the outer
layers of the ejecta, leading to a high opacity consistent with its slow light
curve evolution. Interestingly, the detailed spectral modelling of this event
implied a `rise-time' (the time from SN explosion to maximum light in the
$B$-band) longer than that measured from the optical light curve by $\sim 1$
day.  This suggests the existence of a `dark phase' between the SN explosion and
the emergence of the first photons \citep{piro13a,piro13b}, in turn implying
that the bulk of the radioactive heating from \Nifs, which powers the light
curve, must lie deep in the ejecta.

In this paper we present \textit{HST} observations of the second event in our
cycle 18 programme, the nearby SN Ia SN\,2011fe \citep{nugent11}, located in
M101.  Thanks to the proximity of M101 \citep[$\simeq6.4$\,Mpc,
$\mu=29.04\pm0.19$;][]{shappee11}, SN\,2011fe was the brightest SN\,Ia in the
night sky in the `CCD era', and a comprehensive monitoring campaign carried out
across a broad range of wavelengths will make it the best-ever studied SN Ia. 
Our \textit{HST} data range from the far-UV to the near-infrared (IR), beginning
just 4 days after detection and covering 10 epochs.  We supplement these
\textit{HST} data with spectra taken as part of the PTF follow-up campaign
\citep{parrent12}. We use a well-established Monte Carlo code
\citep[e.g.,][]{luc99,maz00,sau08,hachinger13} to model this time series of
UV/optical spectra. Starting from the outermost regions, we infer the abundance
stratification of SN\,2011fe, developing a density profile that yields a good
fit to the spectra from the UV to the near infrared (IR). We also determine the
rise time of the SN from the modelling to compare with the very early SN
photometry.

An outline of the paper is as follows. In Section~\ref{sec:data} we describe the
\textit{HST} observations and their data reduction, and discuss SN\,2011fe in
the context of other SNe Ia in Section~\ref{sec:gener-prop}. We then introduce
the methods used to model the spectra in Section~\ref{sec:methods}, and present
the models themselves in Section~\ref{sec:models} together with our rise time
measurement. We compute spectra for two standard explosion models and for a
custom-made density structure which yields improved fits to the spectra. The
implications of our results are discussed in Section~\ref{sec:discussion} and we
conclude in Section~\ref{sec:conclusions}.

\section{Observations}
\label{sec:data}

SN\,2011fe was discovered on 2011 August 24.167 (UT) by the PTF using the
Palomar 48-in telescope (P48) in M101, at a magnitude of $g\simeq17.4$
\citep[][hereafter N11]{nugent11}. Within a few hours of discovery, the SN was
confirmed as a very early SN Ia using robotic observations with FRODOSPEC on the
Liverpool Telescope \citep[LT;][]{ste04}.  UV/optical/IR spectroscopic
observations were subsequently obtained on 10 epochs with the \textit{Hubble
Space Telescope (HST)} using the Space Telescope Imaging Spectrograph (STIS), as
part of the cycle 18 program \#12298: `Towards a Physical Understanding of the
Diversity of Type Ia Supernovae' (PI: Ellis). The phase 2 was submitted in two
parts: on 2011 August 25 for the the first few epochs, and on 2011 August 29 for
the epochs from maximum light onwards. The later epochs also included parallel
observations with the Wide Field Camera 3 to observe Cepheid variable stars in
M101.

In this section, we describe these \textit{HST} observations and their
reduction, and detail the construction of a single spectrum on each \textit{HST}
epoch. One of these spectra has already been shown in \citet{foley13a}.

\subsection{\textit{HST} spectral observations}
\label{sec:observations}

We used five different STIS configurations using both the CCD and the UV MAMA
detectors, covering 4 different wavelength ranges. On all epochs, the near-UV
was covered using either NUV-MAMA/230L (giving useful coverage from
$\lambda\simeq1750$ to $\simeq3150\textrm{\AA}$) or CCD/230LB
($\lambda\simeq1900$ to $\simeq3050\textrm{\AA}$), the optical using CCD/430L
($\lambda\simeq2950$ to $\simeq5700\textrm{\AA}$), and the near-IR using
CCD/750L ($\lambda\simeq5300\textrm{\AA}$ to $\simeq1\micron$). On one further
epoch near maximum light an additional configuration was used to examine the
far-UV (FUV-MAMA/140L $\lambda\simeq1300$ to $\simeq1700$\AA). When using the
CCD, we split the observations to assist with cosmic ray rejection (`CR-SPLIT'),
and took separate flat-fields after the CCD/750L observations to assist with
fringe removal. We used the 0.2$\arcsec$ slit throughout.  A full log of our
\textit{HST} observations can be found in Table~\ref{tab:speclog}.

At the time of triggering the \textit{HST} ToO, the eventual photometric
evolution of SN\,2011fe, including its peak brightness and light curve width,
was quite uncertain. We thus opted to use the CCD/230LB for the near
maximum-light epochs in preference to NUV-MAMA/230L to avoid any possibility of
exceeding the MAMA bright-object limits, even though the NUV-MAMA/230L was
preferred scientifically because of its superior far-UV coverage. Exposure
times, dithers and CR-SPLITs were chosen in an attempt to avoid saturation of
this bright target while still efficiently filling the time available in each
orbit. For the most part this was successful, although some observations were
slightly saturated at the peaks of the SN spectral features because of an
under-estimate of the target's eventual brightness and photometric evolution. We
discuss the treatment of these saturated spectra in the next section.

\begin{table*}
\caption{Log of the \textit{HST} spectroscopic observations of SN\,2011fe.}
\label{tab:speclog}
\centering
\begin{tabular}{ccccl}
Epoch & MJD observation & Instrument & Total exposure &Comments\\
&(days)&configuration&time (s)&\\
\hline
2011-08-28 & 55801.212 & NUV-MAMA/230L & 5400&\\
2011-08-28 & 55801.100 & CCD/430L & 1200& CR-SPLIT=3\\
2011-08-28 & 55801.117 & CCD/750L & 1002& CR-SPLIT=3\\
\hline
2011-08-31 & 55804.304 & NUV-MAMA/230L & 2200&\\
2011-08-31 & 55804.231 & CCD/430L  & 1156& Dither 2x578s w/ CR-SPLIT=2. Saturation.\\
2011-08-31 & 55804.244 & CCD/750L  & 410& CR-SPLIT=2. Slight saturation.\\
\hline
2011-09-03 & 55807.425 & CCD/230LB & 1320&CR-SPLIT=4.\\
2011-09-03 & 55807.438 & CCD/430L  & 195&CR-SPLIT=3. Slight saturation.\\
2011-09-03 & 55807.444 & CCD/750L  & 195&CR-SPLIT=3.\\
\hline
2011-09-07 & 55811.415 & CCD/230LB  & 1060&Dither 2x530s w/ CR-SPLIT=2.\\
2011-09-07 & 55811.427 & CCD/430L   & 160&Dither 2x80s w/ CR-SPLIT=2.\\
2011-09-07 & 55811.434 & CCD/750L   & 80&CR-SPLIT=2.\\
\hline
2011-09-10 & 55814.417 & CCD/230LB  & 1060&Dither 2x530s w/ CR-SPLIT=2.\\
2011-09-10 & 55814.429 & CCD/430L   & 160&Dither 2x80s w/ CR-SPLIT=2. Slight saturation\\
2011-09-10 & 55814.465 & CCD/750L   & 80&CR-SPLIT=2.\\
\hline
2011-09-13 & 55817.805 & FUV-MAMA/140L & 8540&\\
2011-09-13 & 55817.669 & CCD/230LB  & 830&Dither 2x415s w/ CR-SPLIT=2.\\
2011-09-13 & 55817.680 & CCD/430L   & 160&Dither 2x80s w/ CR-SPLIT=2.\\
2011-09-13 & 55817.689 & CCD/750L   & 140&Dither 2x70s w/ CR-SPLIT=2.\\
\hline
2011-09-19 & 55823.661 & NUV-MAMA/230L & 2720&\\
2011-09-19 & 55823.601 & CCD/430L   & 160&Dither 2x80s w/ CR-SPLIT=2.\\
2011-09-19 & 55823.608 & CCD/750L   & 70&CR-SPLIT=2\\
\hline
2011-10-01 & 55835.298 & NUV-MAMA/230L & 2700&\\
2011-10-01 & 55835.231 & CCD/430L  & 1200&Dither 2x600s w/ CR-SPLIT=3. Slight saturation.\\
2011-10-01 & 55835.244 & CCD/750L  & 190&CR-SPLIT=2.\\
\hline
2011-10-07 & 55841.349 & NUV-MAMA/230L & 3000&Dither 2x1500s\\
2011-10-07 & 55841.281 & CCD/430L  & 1320&Dither 2x660s w/ CR-SPIT=3. Slight saturation.\\
2011-10-07 & 55841.295 & CCD/750L  & 310&CR-SPLIT=2. Slight saturation.\\
\hline
2011-10-21 & 55855.210 & NUV-MAMA/230L & 3000&\\
2011-10-21 & 55855.154 & CCD/430L  & 1220&Dither 2x610s w/ CR-SPLIT=2.\\
2011-10-21 & 55855.168 & CCD/750L  & 420&CR-SPLIT=2.\\
\hline
\end{tabular}
\end{table*}

\subsection{Data reduction}
\label{sec:data-reduction}

The reduction of the \textit{HST} data followed standard procedures.
The data were downloaded from the \textit{HST} archive using the
on-the-fly reprocessing (OTFR) pipeline to provide the appropriate
`best' calibration files. As some aspects of the reduction needed to
be performed `by hand' (e.g. fringe removal in the IR data), we used a
local \textsc{calstis} installation in
\textsc{iraf}\footnote{\textsc{iraf} is distributed by the National
  Optical Astronomy Observatories, which are operated by the
  Association of Universities for Research in Astronomy, Inc., under
  cooperative agreement with the National Science Foundation.} to
perform the reductions.

\textsc{calstis} gives fully calibrated and extracted 1-D spectra, where the
reduction and extraction is optimised for point sources.  It performs initial
2-D image reduction such as image trimming, bias and dark current subtraction,
cosmic-ray rejection (using a CR-SPLIT), and flat-fielding. It then performs 1-D
spectral extraction, followed by wavelength and flux calibrations. In addition
to these standard procedures, we performed three additional steps.

The first step was the fringe frame removal for CCD/750L observations, matching
fringes in the CCD/750L fringe flat-field taken with the science observations,
to those in the science observations themselves. This generally proceeded
satisfactorily, except for the 2011 September 13 epoch where the SN trace was
offset from its standard position to allow a different set of Cepheids to be
observed in the parallel observations.  Unfortunately the fringe-flat position
was not updated to reflect this, hence this epoch has less \textit{HST} near-IR
coverage and a ground-based Gemini IR spectrum is used in its place
\citep{hsiao13}.

The second step was the removal of hot pixels, bad pixels, and residual cosmic
rays in the data. These were identified by hand by examining the post CR-SPLIT
2-D spectral images. These image defects were then interpolated over in the
dispersion direction.

The final additional step was the fixing of the saturation in some of our
optical (CCD/430L and CCD/750L) data at the peaks of the SN spectral features
(see Table~\ref{tab:speclog}). As a point source, the SN has a spatial profile
of $\sim11$ pixels on the CCD, which is very well aligned with the CCD rows.
Where present, saturation affected only the central row of this profile. We
corrected for the saturation by taking the flux ratio of the central row to four
other rows in the profile (two rows above and two rows below the saturated row)
at every wavelength pixel. We then fitted a polynomial function to each of these
four flux ratios as a function of wavelength pixel, using only the unsaturated
regions in the fit. These polynomial fits can then be interpolated across
wavelength pixel regions (columns) that have saturated data, and the SN flux in
the saturated central row of the profile inferred from the interpolated flux
ratios.  The final SN flux in the saturated row is then the weighted mean of
these four values, with an error reflecting the variance in these predictions.
By testing this procedure on unsaturated regions of the SN spectral profile, and
comparing the interpolated flux to the actual observed one, we find an excellent
level of agreement.

\subsection{Combined spectra}
\label{sec:combined-spectra}

The \textit{HST} spectra were extracted using the standard \textsc{calstis} 1-D
spectral extraction tasks, and combined with near-IR data from \citet{hsiao13}
where available to form a single contiguous spectrum on each epoch. Where the
\textit{HST} observations were taken at more than one dither position in a given
configuration, we extract each dither position separately and then combine the
1-D spectra. This is in preference to combining the 2-D images directly, as that
process is sensitive to the exact sub-pixel centering of the spatial profile on
a given row. The individual exposures were rebinned to a common wavelength scale
in each wavelength region: 1.6\AA\ over 1000--2900\AA\ (1.4\AA\ when using the
CCD/230LB), 2.8\AA\ over 2900--5000\AA, 4.9\AA\ over 5000\AA--1.1\micron, and
7\AA\ over 1.1-2.5\micron. The spectra were then optimally combined, weighting
by the flux errors in each input pixel.

We used overlapping regions between the spectra to ensure that the flux
calibration was consistent, adopting the CCD/430L spectrum as the flux reference
at each epoch. Typically, this was a 1--2 per cent correction, although for the
2011-09-13 epoch there was no useful overlap between the FUV-MAMA/140L and the
CCD/230LB data, so we did not adjust the FUV-MAMA/140L flux scale. In all cases
the Gemini near-IR spectra over-lapped with the \textit{HST} CCD/750L spectra to
allow a relative flux calibration between the two.  Here the corrections were
$\sim4-5$ per cent, but larger on the early epochs when the SN flux was
brightening rapidly.  Information on the final mean spectra can be found in
Table~\ref{tab:meanspec}.

\begin{table}
\caption{Details of the mean spectra constructed on each epoch.}
\label{tab:meanspec}
\centering
\begin{tabular}{cccl}
\hline
Epoch & Ave. MJD & Phase & Wavelength \\
&observation& (days)$^{a}$ & coverage\\
\hline
2011-08-28 & 55801.17 & -13.1 & 1750\AA\ -- 2.5$\micron$\\
2011-08-31 & 55804.25 & -10.1 & 1750\AA\ -- 2.5$\micron$\\
2011-09-03 & 55807.38 & -6.9  & 1900\AA\ -- 2.5$\micron$\\
2011-09-07 & 55811.37 & -2.9  & 1900\AA\ -- 2.5$\micron$\\
2011-09-10 & 55814.39 & +0.1  & 1900\AA\ -- 2.5$\micron$\\
2011-09-13 & 55817.67 & +3.4  & 1265\AA\ -- 2.5$\micron$\\
2011-09-19 & 55823.62 & +9.3  & 1750\AA\ -- 1.02$\micron$\\
2011-10-01 & 55835.26 & +20.9 & 1750\AA\ -- 1.02$\micron$\\
2011-10-07 & 55841.31 & +27.0 & 1750\AA\ -- 1.02$\micron$\\
2011-10-21 & 55855.18 & +40.8 & 1750\AA\ -- 1.02$\micron$\\
\hline
\end{tabular}
\parbox{7.7cm}{$^a$The phase is given in days in the SN rest frame relative to 
maximum light in the rest-frame $B$-band.}
\end{table}

The data were corrected for the recession velocity of M101. For M101 as a whole,
this is 241 km\,s$^{-1}$ \citep{1991rc3..book.....D}, but the projected radial
velocity at the SN site is $\simeq180$\,km\,s$^{-1}$
\citep{1981A&A....93..106B,patat13}. We used this latter value in our analysis.
We correct for Milky Way (MW) extinction using the maps of \citet*{sch98} and a
\citet*{car89} extinction law.  Although it is not clear that these dust maps
are reliable near extended objects such as M101, the correction is small (a
colour excess $E(B-V)_{\textrm{mw}} = 0.009$\,mag) and is consistent with
independent evidence from Galactic Na\,\textsc{i} and Ca\,\textsc{ii} absorption
features in high-resolution optical spectra of SN\,2011fe \citep[see][who derive
$E(B-V)_{\textrm{mw}}=0.01$\,mag]{patat13}. We discuss extinction caused by dust
in the host galaxy in the next section.

\begin{figure*}
 \includegraphics[width=160mm]{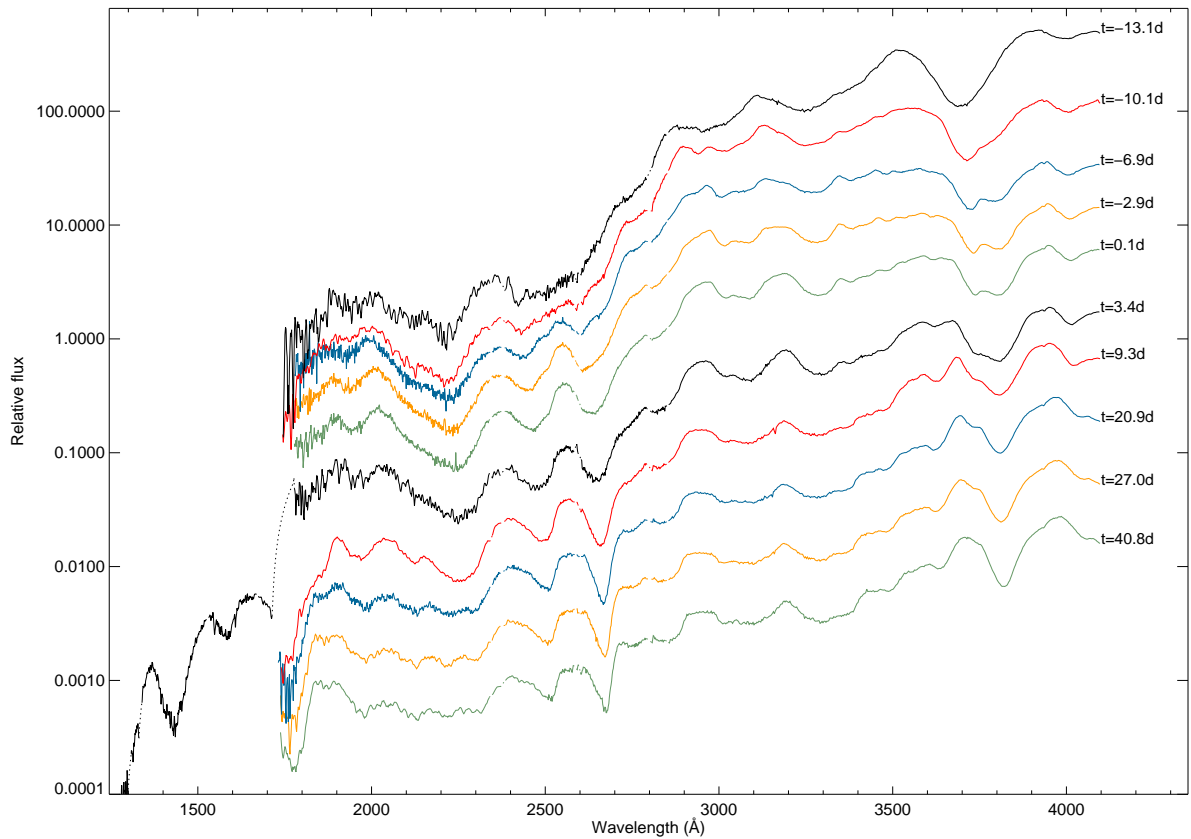}
\caption{A time-series spectral sequence of the UV portion of the
  \textit{HST} spectra of SN\,2011fe.  Spectra have been corrected for
  Milky Way extinction, corrected to the rest-frame, and had UV ISM
  absorption lines removed. The phase of each spectrum, marked in days
  relative to maximum light in the rest-frame $B$-band, is reported on
  the right. Each spectrum has been offset arbitrarily for presentation
  purposes.}
\label{fig:allspec}
\end{figure*}

Some narrow absorption features, presumably from material in the inter-stellar
medium (ISM) along the line of sight to the SN, are present in the UV spectra of
SN\,2011fe, in particular Mg\,\textsc{ii} at 2796\AA\ and 2803\AA,
Mg\,\textsc{i} at 2853\AA, Fe\,\textsc{ii} at 2344\AA, 2374\AA, 2383\AA,
2587\AA\ and 2600\AA, and various Si\,\textsc{ii} and Al\,\textsc{ii} features
in the far-UV. These are at velocities that are consistent with the recession
velocity of M101 given the STIS resolution and wavelength calibration accuracy,
and are probably associated with the system causing strong Na\,\textsc{i}
absorption at 180 km\,s$^{-1}$ identified by \citet{patat13}. We remove the
strongest of these lines from the spectra in this paper.  The spectra are shown
in Fig~\ref{fig:allspec}, and are available from the WISeREP
archive\footnote{\href{http://www.weizmann.ac.il/astrophysics/wiserep/}{http://www.weizmann.ac.il/astrophysics/wiserep/}}
\citep{yaron12}.

\section{General properties of SN\,2011fe}
\label{sec:gener-prop}

We now discuss the general optical and UV properties of SN\,2011fe. In
particular, we assess whether SN\,2011fe is broadly representative of
other SNe Ia in the UV.

\subsection{Photometric properties}
\label{sec:phot-prop}

M101 was being observed by PTF with a daily cadence around the time of the
SN\,2011fe discovery. There are non-detection limits of $g>21.2$ on MJD 55795.2
and $g>22.2$ on 55796.2, followed by the first detection at $g=17.3\pm0.01$ on
MJD 55797.2. The explosion date of SN\,2011fe was estimated by
\citetalias{nugent11} by fitting a simple power-law model to the early-time P48
SN flux, $f$, as a function of time $t$, given by
$f(t)\propto(t-t_{\textrm{expl}})^{n}$, where $t_{\textrm{expl}}$ is the
explosion time and $n$ is an exponent: $n=2$ corresponds to a simple `fireball'
model. \citetalias{nugent11} find $t_{\textrm{expl}}=55796.687\pm0.014$ in MJD
(or 2011 August 23.69), with an exponent $n=2.01\pm0.01$.

A significant amount of additional optical photometry of SN\,2011fe exists
\citep{richmond12,vinko12,munari13}, and is compared in detail by
\citet{pereira13}. Here, we use the SiFTO light curve fitter \citep{conley08} to
fit $g$ photometry from the P48 (as this provides the earliest photometric
coverage) and $BVR$ data published by \citet{vinko12}.  We estimate a time of
maximum light in the rest-frame $B$-band of MJD $55814.30\pm0.06$ (2011
September 10.3), and a peak magnitude in the $B$-band of $9.93\pm0.02$\,mag. The
phases of the mean spectra on each epoch can be found in
Table~\ref{tab:meanspec}.  SN\,2011fe has a SiFTO light curve stretch of
$0.98\pm0.01$, and from the SiFTO templates we estimate $\Dm=1.05$\,mag --
although we caution this number is not a product of SiFTO and is an estimate
only. Similar values have been measured by other authors
\citep[e.g.][]{munari13}. The $B-V$ colour, \col, at $B$-band maximum light is
$\col=-0.07\pm0.02$.  These fit parameters are all fully consistent with
independent SALT-II \citep{guy07} fits to SNfactory spectrophotometric data
presented by \citet{pereira13} and, as discussed by those authors, make
SN\,2011fe a typical example of a normal SN Ia, and one that would be included
in any standard Hubble diagram analysis had it been located in the Hubble flow
\citep[see][for a summary of the criteria typically used]{conley11}.

The $B-V$ colour does not provide any evidence for extinction due to the host
galaxy; if anything it is slightly blue \citep[cf. figure 1 of][]{conley11}.
Independent evidence for low extinction comes from \citet{patat13}, who measure
$E(B-V)_{\textrm{host}}=0.014\pm0.002$\,mag. We do not correct the data for this
small extinction but do account for it in our modelling.

\begin{figure*}
\includegraphics[width=80mm]{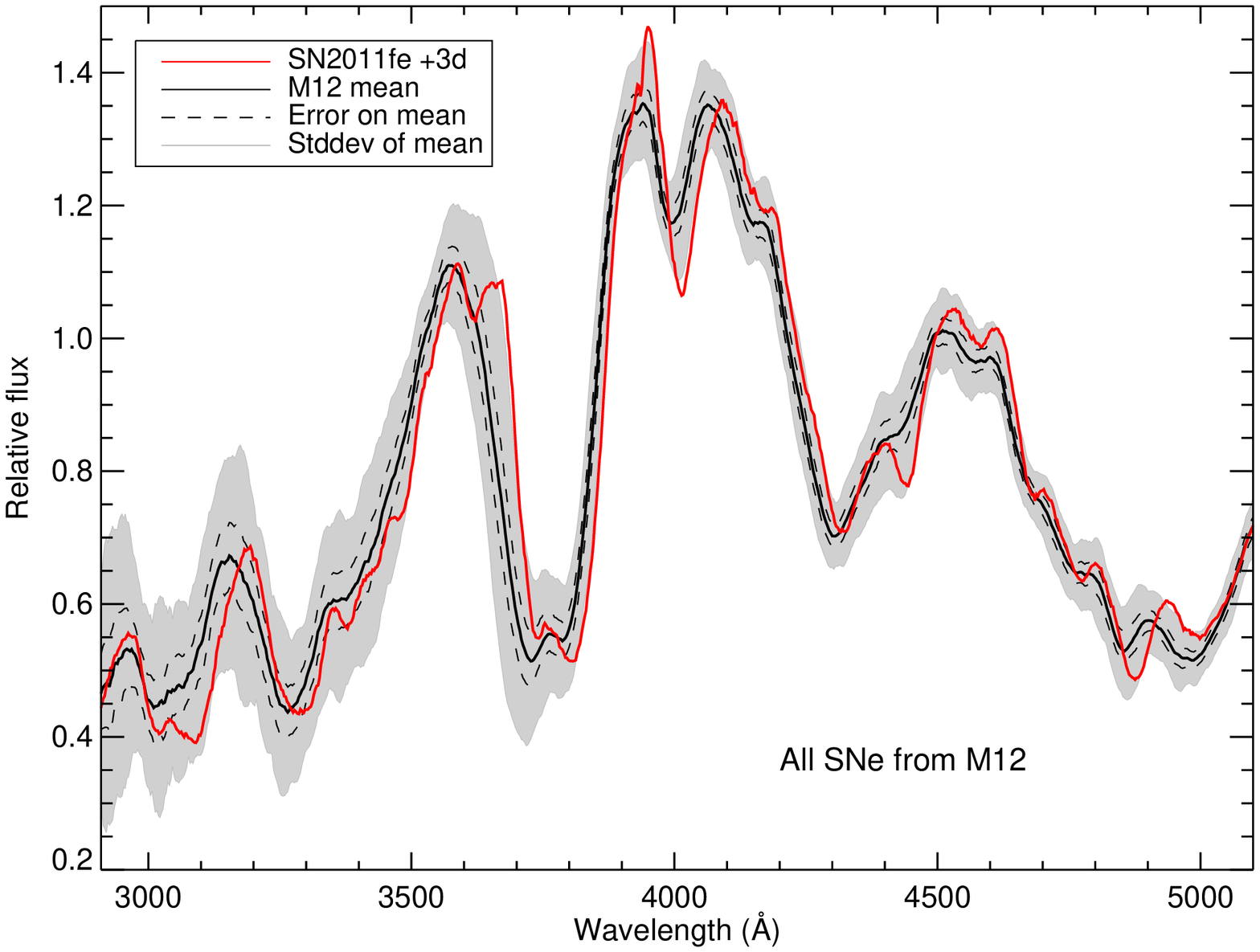}
\includegraphics[width=80mm]{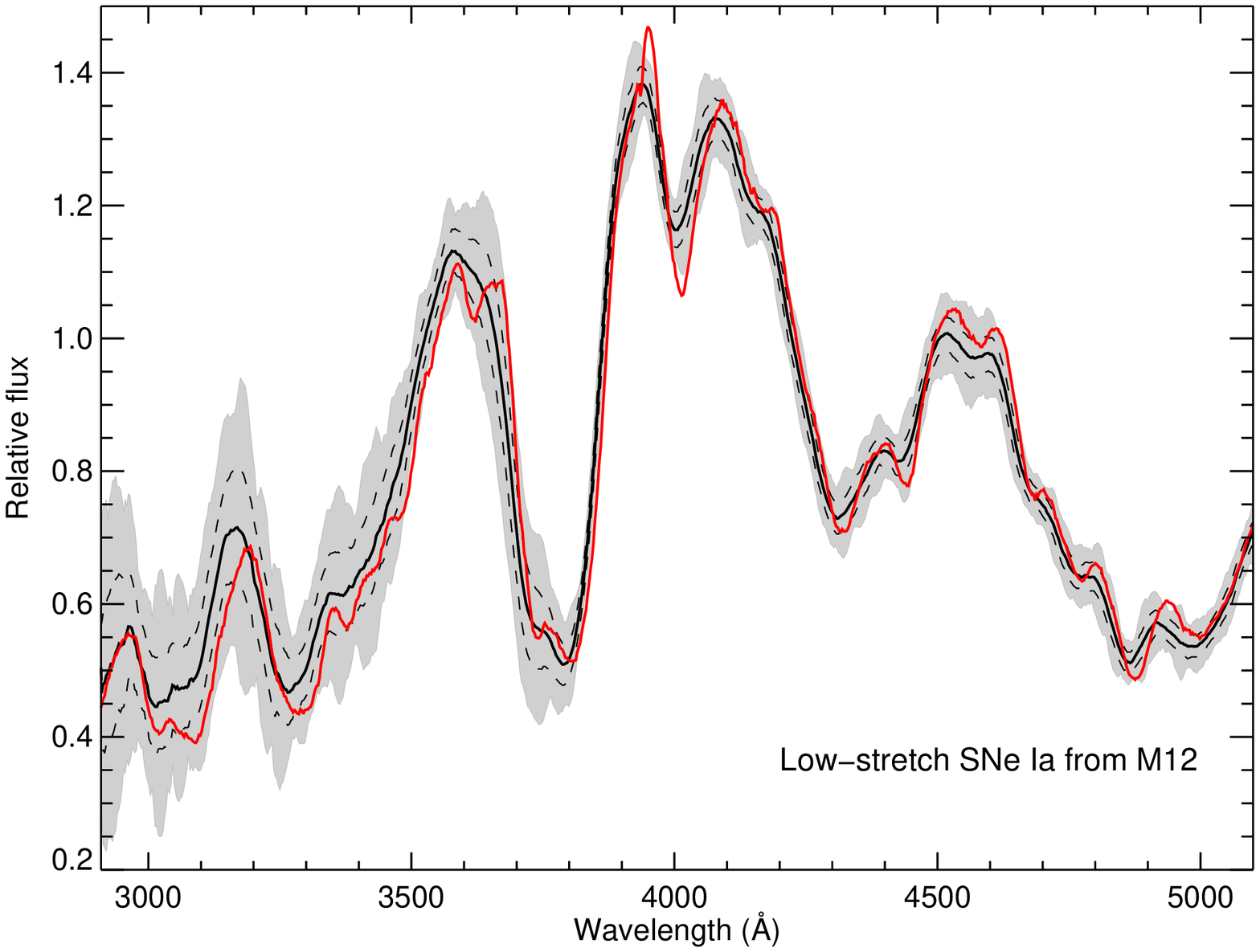}
\caption{A comparison between the average of the +0.1\,d and +3.4\,d
  spectra of SN\,2011fe, and the mean near-UV spectrum of
  \citetalias{maguire12} constructed from \textit{HST} data taken at phases 
  between -1\, and +4.5\,d, excluding SN\,2011fe. The dashed line is the error 
  on the \citetalias{maguire12} mean spectrum, and the shaded grey band shows
  the standard deviation of the input data. The spectra have been
  normalised using a box filter between 4000\AA\ and 4500\AA.  
  Left: The comparison to the entire \citetalias{maguire12} sample.  
  Right: The comparison to just the low stretch events ($s<1.03$).}
\label{fig:maguirecmp}
\end{figure*}

The inferred rise time \tr\ measured from the SN light curve is $17.6\pm0.1$\,d,
close to the average for normal SNe Ia \citep{hayden10a}. This is most likely a
lower limit to the true \tr, i.e., the true explosion date is likely earlier
than that measured from the light curve: the light-curve based \tr\ measurement
does not take into account photon diffusion within the dense ejecta of the young
SN.  This can lead to a delay between the SN explosion and the emergence of the
first optical photons \citep{piro13a,piro13b,hachinger13}. The rise-time (or
explosion date) is a critical input to our modelling as it defines the time from
the explosion of the SN to each of our spectra. We assess the consistency of
various values of \tr\ with the spectroscopic modelling in
Section~\ref{sec:models-risetime-iwamoto}.

\label{sec:spec-comp}

\subsection{UV spectral comparisons}

We compare our near-UV spectra of SN\,2011fe with the mean near-UV spectrum for
low redshift SNe Ia compiled from data in \citetalias{maguire12} from -1 to
+4.5\,d. Two of our spectra have phases in this range (Table~\ref{tab:meanspec})
and we average them together to give a mean phase of +1.8\,d. The
\citetalias{maguire12} mean includes SN\,2011fe in its construction, so we
re-generate the mean without SN\,2011fe; the average phase of the mean spectrum
is +2.1\,d.  The comparison is shown in Fig.~\ref{fig:maguirecmp} (left-hand
panel).  The overall level of agreement is good, although some differences can
be seen: for example, the positions/velocities of the near-UV features are
systematically redder/slower in SN\,2011fe compared to the mean of the
\citetalias{maguire12} sample.

\citetalias{maguire12} identified a strong trend between the stretch of SNe Ia
and the velocities/positions of some near-UV spectral features; higher stretch
SNe Ia have bluer features. As SN\,2011fe, with $s=0.98$
(Section~\ref{sec:phot-prop}) would fall into the `low stretch' sample of
\citetalias{maguire12}, we also compare to the mean generated from just the
low-stretch SNe (Fig.~\ref{fig:maguirecmp}, right-hand panel).  The agreement
between the UV feature positions (particularly the Ca\,\textsc{ii} H\&K
absorption) is better, and the SN\,2011fe spectrum falls within the range of
spectra examined in \citetalias{maguire12}.

In summary to this section, we conclude, as have others, that SN\,2011fe is
photometrically a fairly typical example of a SN Ia, and that this extends to
its behaviour in the near-UV. Thus inferences drawn from SN\,2011fe are likely
to be broadly applicable to other SNe Ia.

\section{Modelling technique}
\label{sec:methods}

We now turn to the modelling of the SN\,2011fe spectra. We first
outline our modelling framework and assumptions, and then present the
models of SN\,2011fe in Section~\ref{sec:models}.

\subsection{Monte Carlo radiative transfer code}
\label{sec:mont-radi-transf}

We model the spectra of SN\,2011fe using a Monte Carlo code
\citep{abb85,mazzali93b,luc99,maz00,ste05}. The code performs radiative transfer
above a sharp lower boundary (or `photosphere'). To generate a spectrum at any
given epoch, the code requires as input the bolometric luminosity \lbol, the
time $t$ since the SN explosion (hence the importance of an accurate knowledge
of \tr), the photospheric velocity $v_\textrm{ph}$, the density distribution of
the SN ejecta in the homologous expansion phase
(Section~\ref{sec:models-densityprof}), and the abundances of the elements in
the ejecta.

The code simulates the propagation of photon packets emitted at the photosphere
with a black-body spectrum of temperature $T_{\textrm{ph}}$ [$I_{\nu}^{+}$\, =
\,$B_{\nu} (T_{\textrm{ph}})$].  The SN ejecta are assumed to be optically thick
below this photosphere. Although this is a relatively rough approximation, and
one which at later epochs can cause an excess in the red and IR flux of the
models, it makes the code flexible as quantitative abundances can be derived
without a knowledge of the exact distribution of radioactive heating below the
photosphere. Since all lines important for abundance determination are in the
UV/optical, the poor reproduction of the near-IR continuum has only minor
consequences on our results.  Photon packets undergo Thomson scattering and line
absorption.  Following absorption, packets are immediately re-emitted in a
transition chosen randomly via a branching scheme. Packets that are scattered
back into the photosphere are considered to be re-absorbed. Radiative
equilibrium is enforced by the `indivisible-packet' approach \citep{luc99}. The
code iterates $T_{\textrm{ph}}$ (and thus the outgoing intensity at the
photosphere) so as to match \lbol\ given the actual back-scattering rate.

The excitation/ionisation state of the gas is determined by the
radiation field using a modified nebular approximation
\citep{mazzali93b,maz00}. Starting from an initial guess, the
radiation field and the state of the gas are iterated with a series of
Monte Carlo experiments until convergence is achieved. The final
emergent spectrum is obtained by computing the formal integral,
avoiding excessive Poisson noise \citep{luc99}.

\subsection{Spectral modelling}
\label{sec:spectral-modelling}

In its simplest form, our code uses homogeneous abundances above the photosphere
\citep[`one-zone' modelling, e.g.][]{mazzali93a}. Intermediate-mass elements
(IMEs, e.g. Mg, Si, S, Ca) typically influence the strength of spectral features
in the optical, whereas Fe-group elements lead to the formation of features in
the optical and the UV.  Additionally, Fe-group elements are essential for
fluorescence and reverse-fluorescence processes, as simulated by the branching
scheme \citep{luc99}. These processes can be particularly important for the
formation of the UV spectrum \citep{maz00}.

The spectral line strengths are strongly influenced by ionisation/excitation,
which in turn depend on the radiation field.  A lower $v_\textrm{ph} (=
r_\textrm{ph}/t)$ leads to a lower $r_\textrm{ph}$ and hence a higher
temperature with a bluer radiation field at the photosphere, favouring
ionisation and the occupation of highly-excited states, with lines forming at
lower velocity. A higher \lbol\ has a similar effect on the temperature but not
on the line velocities, apart from an indirect effect on ionisation and
excitation.

The typical modelling process involves iterating \lbol\ until the flux in the
observed spectrum is matched by the model; any IR excess in the model due to the
photospheric black-body approximation (Section~\ref{sec:mont-radi-transf}) is
disregarded. Simultaneously, $v_\textrm{ph}$ is iterated to optimise the
position of the lines and the overall spectral shape.  Abundances are first
defined in an approximate way during this process, and may be improved once
\lbol\ and $v_\textrm{ph}$ are fixed, but since they both depend on the density
structure and affect the thermal state of the gas, the two procedures are not
physically independent. This also implies that it may be difficult to find a
`perfect' solution. However, the combined fit of line position and strength with
the shape of the pseudo-continuum represents a quantitative solution of the
physical conditions of the SN atmosphere.

\subsection{Abundance tomography}
\label{sec:abundance-tomography}

Although one-zone modelling can effectively constrain abundances near the
photosphere, it does not follow abundance variations with radius (or velocity).
\citet{ste05} introduced the technique of constraining the abundance
stratification of the ejecta (`abundance tomography') by using a time-series of
SN spectra. Defining the abundances as a function of ejecta depth, they used a
time-series of spectra to `observe' the different layers.

In the early, photospheric phase (up to a few weeks after maximum light in SNe
Ia), the inner ejecta are optically thick.  As the ejecta expand, the
photosphere recedes inside the ejecta following the decreasing density, as
witnessed by the decreasing blueshift of the observed spectral lines
\citep[e.g.][]{ben05}.  In order to obtain an optimal description of the
abundance stratification in the outermost ejecta, two or three independent
layers are usually introduced above the photosphere of the first spectrum. The
abundances in these layers are defined by fitting the spectrum, with the edges
of the layers chosen to improve the fit.

In progressively later spectra the photosphere lies at increasingly lower
velocities. The abundances above the new, lower-lying photosphere (but beneath
the previous one) can now be determined.  This procedure is repeated with the
later spectra until the abundance stratification of the layers accessible with
photospheric-epoch spectra is fully described. The process requires iteration:
abundances in the outer layers determined from earlier spectra may not be
optimal for the later spectra. Thus, the abundances are adapted so as to fit all
spectra in the best possible way, re-calculating the entire spectral sequence to
obtain a consistent model. The uncertainty in this procedure can be as much as
$\sim25$ per cent on the abundances, but it is usually smaller
\citep{mazzali08a}.

The inner layers can only be analysed using nebular spectra, which are needed
to obtain a complete picture and conclude the abundance tomography procedure
\citep{ste05}. The nebular spectra of SN\,2011fe will be analysed in a future
article.

\subsection{Density profiles}
\label{sec:models-densityprof}

One of the most critical ingredients in this process is the assumed density
profile of the SN ejecta, which itself is the product of a particular explosion
model. For example, a density profile with higher densities in the outer layers
of the ejecta can lead to enhanced absorption in the spectral regions sensitive
to these layers. The UV is a particularly sensitive region because the large
number of overlapping metal lines causes the opacity to be higher than in the
optical.  Accordingly, we tested various profiles from different explosion
models on SN\,2011fe. We started with models W7 and WS15DD1
\citep{nom84w7,iwa99}, as they appear to be the most appropriate for SN\,2011fe,
which does not show strong absorption at high velocities. These density profiles
formally correspond to a single-degenerate explosion scenario; we are not able
to test double degenerate scenarios as their density profiles are not available.
However, judging from the plots in \citep{roepke12} the outer layers appear
quite different from those of a delayed detonation in both density and
composition. Other differences may be revealed at
late times \citep[e.g.][]{mazzali11}.

W7 is a parametrised `fast deflagration' model, with kinetic energy $\KE = 1.3
\times 10^{51}$\,erg.  Although it may be considered unphysical because of its
parametrisations, W7 represents a useful benchmark, and its density distribution
has been successfully used to reproduce light curves and spectra of normal SNe
Ia \citep[e.g.][]{ste05,mazzali08a}. Delayed-detonation models, where an initial
deflagration turns into a supersonic detonation at some point during the
explosion \citep{kho91a}, may be considered more realistic. Although the physics
of the transition are still unclear, these models produce explosion energies and
abundance stratifications that match SN Ia observations
\citep{maz07,tanaka11,roepke12}. In particular, they pre-expand the white dwarf
in the deflagration phase, so that the subsequent detonation produces not only
Fe-group elements, but also IMEs, which are ejected at realistic velocities. 
WS15DD1 (hereafter WDD1) is a low-energy delayed-detonation model \citep{iwa99}
with $\KE = 1.33 \times 10^{51}$\,erg and synthesised \Nifs\ mass $M$(\Nifs) =
0.56\,\Msun, which is a good match to values derived from the bolometric
light-curve of SN\,2011fe \citep{pereira13}.

\begin{figure}   
   \centering
   \hspace*{-0.3cm}
   \includegraphics[angle=270,width=8.7cm]{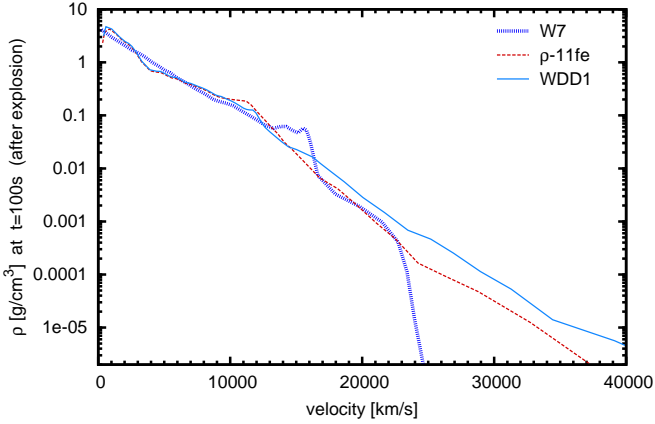}
   \caption{The SN ejecta density profiles of the W7 and WS15DD1
     (WDD1) models discussed in Section~\ref{sec:models-densityprof}
     \citep{iwa99}, and of the $\rho$-11fe model constructed to optimise
     the comparison between the synthetic spectra and observed data
     (Section~\ref{sec:tomography-11fe-w7+}).}
   \label{fig:densitymodels}
\end{figure}

The density profiles of all models used in this paper are shown in
Figure~\ref{fig:densitymodels}. Our models are most sensitive to the density
profile in the outer ejecta, and thus the difference between the amount of
material at high velocity in W7 and WDD1 is important for our models. WDD1 has
more material at high velocity, reflecting the passage of the detonation shock.
Based on the effect density profiles have on the UV, we also built a third,
`intermediate' density profile which yields better fits to the spectra than
either W7 or WDD1. We call this model `$\rho$-11fe'
(Section~\ref{sec:tomography-11fe-w7+}).

\section{Spectral modelling of SN\,2011fe}
\label{sec:models}

Our goal is to model the spectral data for SN\,2011fe using the framework
outlined in Section~\ref{sec:methods}. Our base models assume a metallicity of
0.5\,$Z_{\odot}$, motivated by the metallicity of M101 at the position of
SN\,2011fe \citep{stoll11}. We explore this assumption in
Section~\ref{sec:fe-group-abundances}.

The first step is to determine the rise time (\tr) from the models independently
from the \tr\ measured from the photometry (Section~\ref{sec:phot-prop}). This
requires very early spectral data, where the leverage on the explosion epoch is
largest and the statistical uncertainty consequently smallest \citep{maz00} -- a
small absolute change in \tr\ is large in relative terms. We make use of an
early low-resolution spectrum taken with the Telescopio Nazionale Galileo (TNG)
on 2011 August 25 (MJD 55799.0) and previously presented in
\citetalias{nugent11} and analysed in \citet{parrent12}.  The spectrum of
SN\,2011fe shown here differs from the previously published version as it has
been recalibrated with a more optimum extraction of the spectophotometric
standard star.  This process has made the flux calibration in the blue more
reliable and we now feel more confident about the calibration down to
3500\,\AA.  The redder wavelengths of the spectrum are unaffected.  The value of
\tr\ that provides the best fit is then used to model the spectra at later
epochs.

We did not use two earlier spectra of SN\,2011fe that are also available
\citep[see][]{parrent12}.  The LT and Lick spectra of 2011 August 24 are
affected by high velocity features (HVFs). These are weak compared to other SNe
Ia \citep[e.g.]{mazzali05b} and are not detached from the photospheric
component, but do cause a significant blueshift of the observed features
\citep{parrent12}. This makes it difficult to locate precisely the position of
the photosphere, and hence to produce accurate models, without a detailed
knowledge of the ejecta properties that lead to the generation of these HVFs. We
defer this more complex analysis to a later paper.

\subsection{Testing the rise time}
\label{sec:models-risetime-iwamoto}

\begin{figure*}   
   \centering
   \includegraphics[width=15.0cm,angle=0]{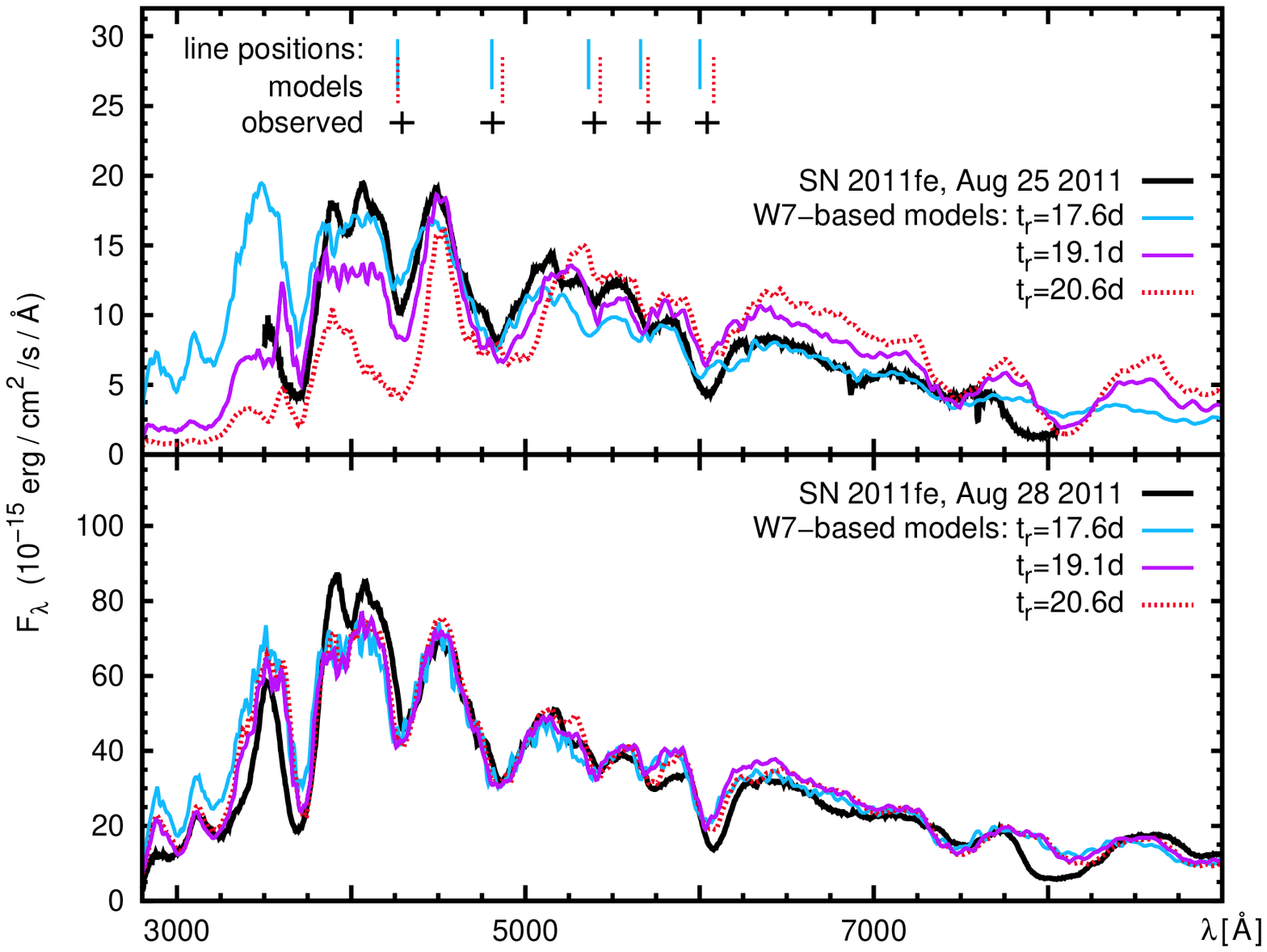}\\[0.6cm]
   \includegraphics[width=15.0cm,angle=0]{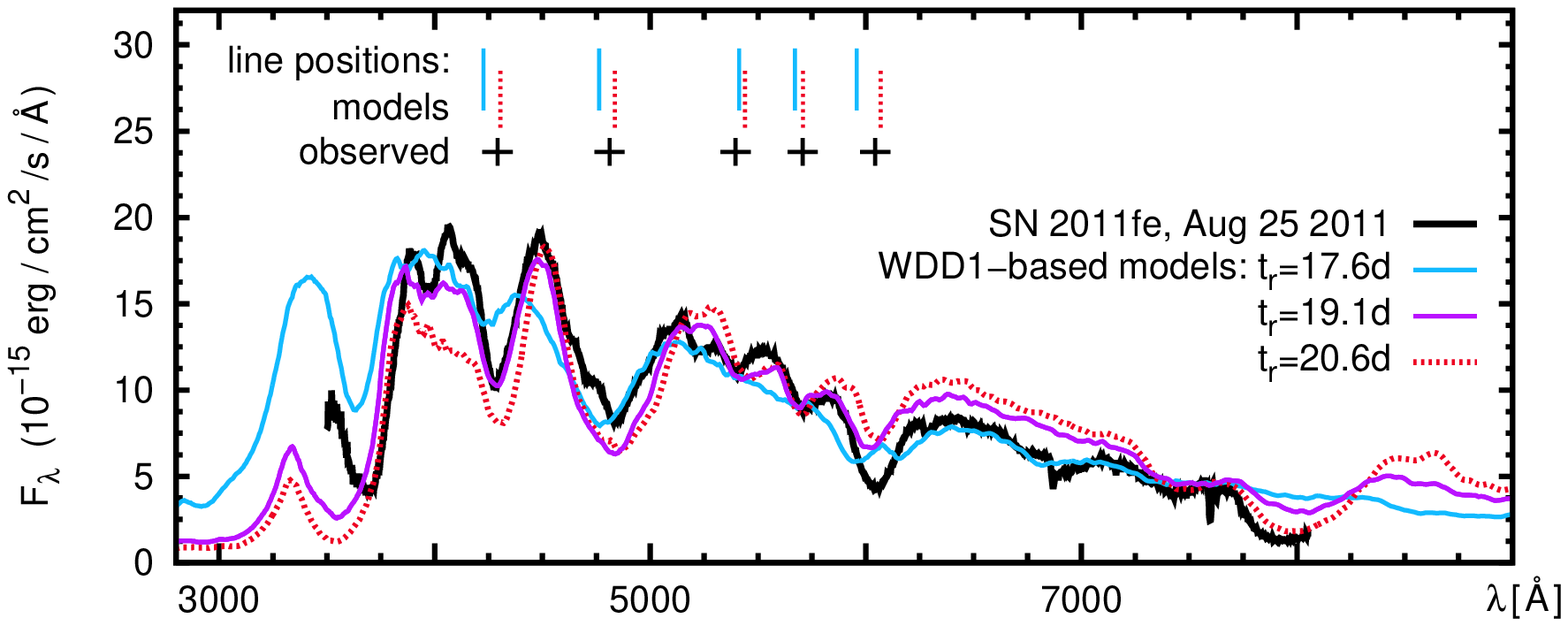}      
   \caption{Early-time models for SN\,2011fe computed for rise times
     \tr\ ranging from 17.6 to 20.6\,d. Models in the \textit{upper
       panel} were computed with the W7 density profile, models in
     the \textit{lower panel} are based on WDD1. The best-fitting \tr\
     is determined from the 2011 August 25 spectrum, where \tr\ has
     the largest effect: the middle panel shows the W7 models for the
     2011 August 28 spectrum, which are less sensitive to \tr.  
     The positions of prominent features in the spectra are marked 
     as black crosses, and the corresponding model features also marked.  
     For \tr=17.6\,d, the model spectral features are on average too blue; 
     this is no longer the case when a longer \tr\ is used.}
   \label{fig:models-11fe-rt-iwamoto}
\end{figure*}

\begin{table*}
  \caption{
    Rise time (\tr) determination from fits to the 2011 August 25
    SN\,2011fe spectrum, with the offset between synthetic and observed spectral features, 
    $\Delta\lambda=\lambda_\textrm{model} - \lambda_\textrm{obs}$ and the average
    relative offset, $\overline{\Delta\lambda}_\textrm{rel}$ (cf.\ text). The fact
    that features do not always follow the overall trend of being less blueshifted
    for larger \tr\ reflects the uncertainties in determining model line velocities.}
\label{tab:linevel-measurements-iwamoto}
\centering
\begin{tabular}{lrrrrrr}
  $t_r$ & 
  $\ \ $ $\Delta\lambda$(Fe\,/\,Mg $\sim$\,4300\,\AA) $\!\!\!\!\!\!$ & 
  $\ \ $ $\Delta\lambda$(Fe\,/\,etc. $\sim$\,4800\,\AA) $\!\!\!\!\!\!$ &
  $\ $ $\Delta\lambda$(\SII\ $\lambda$5640) $\!\!$ &
  $\!\!$ $\Delta\lambda$(\SII\ $\lambda$5972) $\!\!$ &
  $\!\!$ $\Delta\lambda$(\SiII\ $\lambda$6355) $\!\!$ & 
  $\!\!$ $\overline{\Delta\lambda}_\textrm{rel}$ $\!\!\!\!$\\ 
  & $\!\!\!\!$ (\AA) $\qquad$  & $\!\!\!\!$ (\AA) $\qquad$  & $\!\!\!\!$ (\AA)
  $\qquad$  & $\!\!\!\!$ (\AA) $\qquad$  & $\!\!\!\!$
  (\AA) $\qquad$  & $\!\!\!\!$ (\%) $\!\!\!\!$  \\
  \hline
  \multicolumn{7}{c}{W7-based models}\\
  \hline
$\!\!\!\!$	17.6	$\qquad$ & $\!\!\!\!$	$-$25.1	$\qquad$ & $\!\!\!\!$	$-$5.1	$\qquad$ & $\!\!\!\!$	$-$33.8	$\qquad$ & $\!\!\!\!$	$-$46.0	$\qquad$ & $\!\!\!\!$	$-$42.1	$\qquad$ & $\!\!\!\!$	$-$0.56	$\!\!\!\!$	\\
$\!\!\!\!$	18.1	$\qquad$ & $\!\!\!\!$	$-$16.5	$\qquad$ & $\!\!\!\!$	2.6	$\qquad$ & $\!\!\!\!$	$-$19.9	$\qquad$ & $\!\!\!\!$	$-$45.9	$\qquad$ & $\!\!\!\!$	$-$28.4	$\qquad$ & $\!\!\!\!$	$-$0.39	$\!\!\!\!$	\\
$\!\!\!\!$	18.6	$\qquad$ & $\!\!\!\!$	$-$3.6	$\qquad$ & $\!\!\!\!$	10.8	$\qquad$ & $\!\!\!\!$	$-$0.2	$\qquad$ & $\!\!\!\!$	$-$37.8	$\qquad$ & $\!\!\!\!$	$-$18.5	$\qquad$ & $\!\!\!\!$	$-$0.17	$\!\!\!\!$	\\
$\!\!\!\!$	19.1	$\qquad$ & $\!\!\!\!$	10.0	$\qquad$ & $\!\!\!\!$	15.7	$\qquad$ & $\!\!\!\!$	11.8	$\qquad$ & $\!\!\!\!$	$-$35.4	$\qquad$ & $\!\!\!\!$	$-$0.3	$\qquad$ & $\!\!\!\!$	0.03	$\!\!\!\!$	\\
$\!\!\!\!$	19.6	$\qquad$ & $\!\!\!\!$	15.1	$\qquad$ & $\!\!\!\!$	30.6	$\qquad$ & $\!\!\!\!$	26.3	$\qquad$ & $\!\!\!\!$	$-$17.6	$\qquad$ & $\!\!\!\!$	18.0	$\qquad$ & $\!\!\!\!$	0.29	$\!\!\!\!$	\\
$\!\!\!\!$	20.1	$\qquad$ & $\!\!\!\!$	3.6	$\qquad$ & $\!\!\!\!$	47.8	$\qquad$ & $\!\!\!\!$	41.8	$\qquad$ & $\!\!\!\!$	$-$21.1	$\qquad$ & $\!\!\!\!$	22.3	$\qquad$ & $\!\!\!\!$	0.37	$\!\!\!\!$	\\
$\!\!\!\!$	20.6	$\qquad$ & $\!\!\!\!$	$-$23.9	$\qquad$ & $\!\!\!\!$	57.2	$\qquad$ & $\!\!\!\!$	33.6	$\qquad$ & $\!\!\!\!$	$-$2.9	$\qquad$ & $\!\!\!\!$	37.5	$\qquad$ & $\!\!\!\!$	0.37	$\!\!\!\!$	\\
\hline
  \multicolumn{7}{c}{$\!\!\!\!$regression line fit to
$\overline{\Delta\lambda}_\textrm{rel}$--\tr\ relation
$\Rightarrow$ $\overline{\Delta\lambda}_\textrm{rel}\simeq0$ for
   $t_\textrm{r,W7}\simeq\textrm{19.1}$\,d$\!\!\!\!$}\\
  \hline
  \multicolumn{7}{c}{WDD1-based models}\\
  \hline
$\!\!\!\!$	17.6	$\qquad$ & $\!\!\!\!$	$-$53.9	$\qquad$ & $\!\!\!\!$	$-$46.8	$\qquad$ & $\!\!\!\!$	0.5	$\qquad$ & $\!\!\!\!$	$-$52.4	$\qquad$ & $\!\!\!\!$	$-$87.1	$\qquad$ & $\!\!\!\!$	$-$0.92	$\!\!\!\!$	\\
$\!\!\!\!$	18.1	$\qquad$ & $\!\!\!\!$	$-$33.2	$\qquad$ & $\!\!\!\!$	$-$27.2	$\qquad$ & $\!\!\!\!$	$-$16.0	$\qquad$ & $\!\!\!\!$	$-$48.4	$\qquad$ & $\!\!\!\!$	$-$68.1	$\qquad$ & $\!\!\!\!$	$-$0.72	$\!\!\!\!$	\\
$\!\!\!\!$	18.6	$\qquad$ & $\!\!\!\!$	$-$17.9	$\qquad$ & $\!\!\!\!$	$-$18.0	$\qquad$ & $\!\!\!\!$	$-$7.3	$\qquad$ & $\!\!\!\!$	$-$37.1	$\qquad$ & $\!\!\!\!$	$-$44.6	$\qquad$ & $\!\!\!\!$	$-$0.46	$\!\!\!\!$	\\
$\!\!\!\!$	19.1	$\qquad$ & $\!\!\!\!$	$-$2.8	$\qquad$ & $\!\!\!\!$	$-$4.4	$\qquad$ & $\!\!\!\!$	24.3	$\qquad$ & $\!\!\!\!$	$-$26.8	$\qquad$ & $\!\!\!\!$	$-$17.8	$\qquad$ & $\!\!\!\!$	$-$0.09	$\!\!\!\!$	\\
$\!\!\!\!$	19.6	$\qquad$ & $\!\!\!\!$	8.5	$\qquad$ & $\!\!\!\!$	8.7	$\qquad$ & $\!\!\!\!$	45.0	$\qquad$ & $\!\!\!\!$	$-$21.2	$\qquad$ & $\!\!\!\!$	$-$4.6	$\qquad$ & $\!\!\!\!$	0.15	$\!\!\!\!$	\\
$\!\!\!\!$	20.1	$\qquad$ & $\!\!\!\!$	8.8	$\qquad$ & $\!\!\!\!$	15.0	$\qquad$ & $\!\!\!\!$	42.5	$\qquad$ & $\!\!\!\!$	$-$11.2	$\qquad$ & $\!\!\!\!$	13.3	$\qquad$ & $\!\!\!\!$	0.27	$\!\!\!\!$	\\
$\!\!\!\!$	20.6	$\qquad$ & $\!\!\!\!$	14.4	$\qquad$ & $\!\!\!\!$	25.9	$\qquad$ & $\!\!\!\!$	53.5	$\qquad$ & $\!\!\!\!$	$-$1.6	$\qquad$ & $\!\!\!\!$	20.3	$\qquad$ & $\!\!\!\!$	0.43	$\!\!\!\!$	\\
  \hline
  \multicolumn{7}{c}{$\!\!\!\!$regression line fit to
$\overline{\Delta\lambda}_\textrm{rel}$--\tr\
    relation $\Rightarrow$ $\overline{\Delta\lambda}_\textrm{rel}\simeq0$ for
    $t_\textrm{r,WDD1}\simeq\textrm{19.5}$\,d$\!\!\!\!$}\\
  \hline
\end{tabular}
\begin{flushleft}
\end{flushleft}
\end{table*}

Early-time spectra of SNe can be used to estimate the time of the explosion
\citep{mazz_sch05}.  The combination of \lbol, temperature (which depends on
\lbol\ and radius $r$), and velocity of the absorption lines (which depends on
$r$ and $t$) can give powerful constraints on \tr.  For both the W7 and WDD1
profiles, we computed spectral models on epochs ranging from 2.3 to 5.3 d,
corresponding to rise times from 17.6\,d \citepalias{nugent11}, which we regard
as a lower limit, to 20.6\,d, in steps of 0.5\,d.  A selection of these
different \tr\ models is shown in Fig.~\ref{fig:models-11fe-rt-iwamoto}. The
different assumed epochs require different values of $v_\textrm{ph}$. The line
velocities reflect the time evolution of the density profile, and thus in
particular the assumed \tr\ \citep{hachinger12b}.  For a larger \tr, the spectra
will be older, and thus the ejecta density in the model will be lower;
$v_\textrm{ph}$ is therefore lower and the lines less blueshifted. Note that
$v_\textrm{ph}$ is not exactly inversely proportional to $t$, as would be
predicted by a simple Stefan-Boltzmann law. In models with higher $t$ or lower
$v_\textrm{ph}$ (i.e. a photosphere deeper in the density profile, which rises
steeply inwards) the back-scattering rate of photons would be strongly increased
if one assumed $v_\textrm{ph}\sim t^{-1}$. Therefore, a realistic photosphere
for such a model, avoiding an excessive temperature increase at the photosphere
(via strong back-warming to maintain the outwards photon flux), must lie at a
higher velocity.

For the August 25 spectrum, $v_\textrm{ph}$ lies between $\sim10000$ and
18000\,\kms, depending on density profile and \tr.  We chose values that yield
the same radiation temperature at the photosphere ($T_\textrm{ph}$) in all
models with the same density profile: $\sim$10800\,K for models based on W7 and
$\sim$10300\,K for models based on WDD1. This ensures that the
ionisation/excitation state is similar in all models. Two abundance zones were
used above $v_\textrm{ph}$, with the boundaries between these zones at
19400\,\kms\ for W7 and at 18100\,\kms\ for WDD1. Abundances were optimised for
the models with \tr=19.1\,d and not further modified; this has only a minor
influence on the results. 

The optimum \tr\ is taken as the one where the line positions in the
model spectra best match the line positions in the observations. This
is only possible if the lines measured do not exhibit strong HVFs
\citep{mazzali05b}. In the TNG spectrum the HVFs are present only in
the Ca lines, and we exclude these from the determination of \tr.

We measured the positions\footnote{Line-positions are measured as a line
centroid wavelength, calculated as $\lambda_c = \frac{\int_{\Delta\lambda}
\textit{FD}^3(\lambda) \lambda \textrm{d}\lambda}{\int_{\Delta\lambda}
\textit{FD}^3(\lambda) \textrm{d}\lambda}$, where $\textit{FD}^3(\lambda)$ is
the third power of the fractional depth at $\lambda$ \citep[cf.][]{hac08}.
Integrals run over the range $\Delta\lambda$ in which the line absorbs. We use
the centroid instead of the deepest point \citep[as in][]{hac08} as the centroid
can be more reliably determined when feature shapes are irregular and show
considerable fluctuations within the model series, as is the case for our very
early-epoch models here.} of features that are strong in the first spectrum and
well reproduced in the models (marked in Fig.~\ref{fig:models-11fe-rt-iwamoto}),
as in \citet{hachinger12b}. Table~\ref{tab:linevel-measurements-iwamoto} gives
the wavelength offset $\Delta\lambda = \lambda_\textrm{model} -
\lambda_\textrm{obs}$ for the two different models and the spectral features.
Fitting a regression line to the $\overline{\Delta\lambda}_\textrm{rel}$--\tr\
relation (where $\overline{\Delta\lambda}_\textrm{rel}$ is the mean of the
relative wavelength offsets
$\Delta\lambda_\textrm{rel}=\frac{\Delta\lambda}{\lambda_\textrm{obs}}$ of the
features at fixed \tr), we determine the value of \tr\ for which
$\overline{\Delta\lambda}_\textrm{rel}=\textrm{0}$. We obtain an optimum \tr\ of
19.1\,d for W7 and 19.5\,d for WDD1, with a statistical error of $\simeq0.5$\,d
in both cases.  The relation $\overline{\Delta\lambda}_\textrm{rel}$--\tr\ in
Table~\ref{tab:linevel-measurements-iwamoto} is obviously influenced by errors
in the measured feature centroids on a scale $\Delta \tr \lesssim 0.5$\,d (see
e.g. the similar $\overline{\Delta\lambda}_\textrm{rel}$ values for the W7
models with 20.1\,d and 20.6\,d), but on larger time scales it is well-behaved.

The quality of the spectral models depends on the shape of the entire spectrum, 
and not just on the match in the measured wavelength. Spectra computed for a
long \tr\  (20.6\,d) do not match the data. They  are too red, reflecting the
fact that the photosphere is too deep, causing too much absorption.  The WDD1
model for a short \tr\  (17.6\,d) shows features that are too blue and weak, as
the mass above the photosphere is too small. The spectrum computed with the W7
model for \tr\ $=17.6$\,d shows somewhat high velocities in the features, but
otherwise it is compatible with the data, leading to a smaller \tr\ estimate for
W7 than for WDD1. 

Although the optimum \tr\ somewhat depends on the density profile used, and so
some small systematic uncertainty (some 0.1 day) may be expected, it is always
larger than 17.6\,d by $\gtrsim$1\,d. This suggests that
SN\,2011fe exploded earlier than inferred by fitting a power-law model to the
early light curve. This delay is a simple effect of photon diffusion: visible
light is produced by the thermalisation of the $\gamma$-rays and the deposition
of the energy of the positrons emitted in the decay of \Nifs.  At early phases,
when the density is high, these processes take place essentially locally in the
region where \Nifs\ dominates, typically in the deeper layers of the ejecta.
Optical photons must then diffuse out of the ejecta before they can be observed.
This requires time, given the high density and consequently high opacity in the
inner ejecta.

We conclude that while a fit to early light curve data can only set a lower
limit on \tr, with spectral modelling it is possible to improve this estimate.
The accuracy of the result depends on the availability of early spectral data. 
However, in the presence of any HVFs that affect the spectra, standard explosion
models that are unable to reproduce these HVFs make it difficult (if not
impossible) to determine \tr\ without a detailed modelling of the outermost
layers of the SN ejecta \citep{tan08}.

\subsection{W7- and WDD1-based spectral models}
\label{sec:tomography-11fe-iwamoto}

\begin{figure*}   
   \centering
   \includegraphics[width=15.2cm]{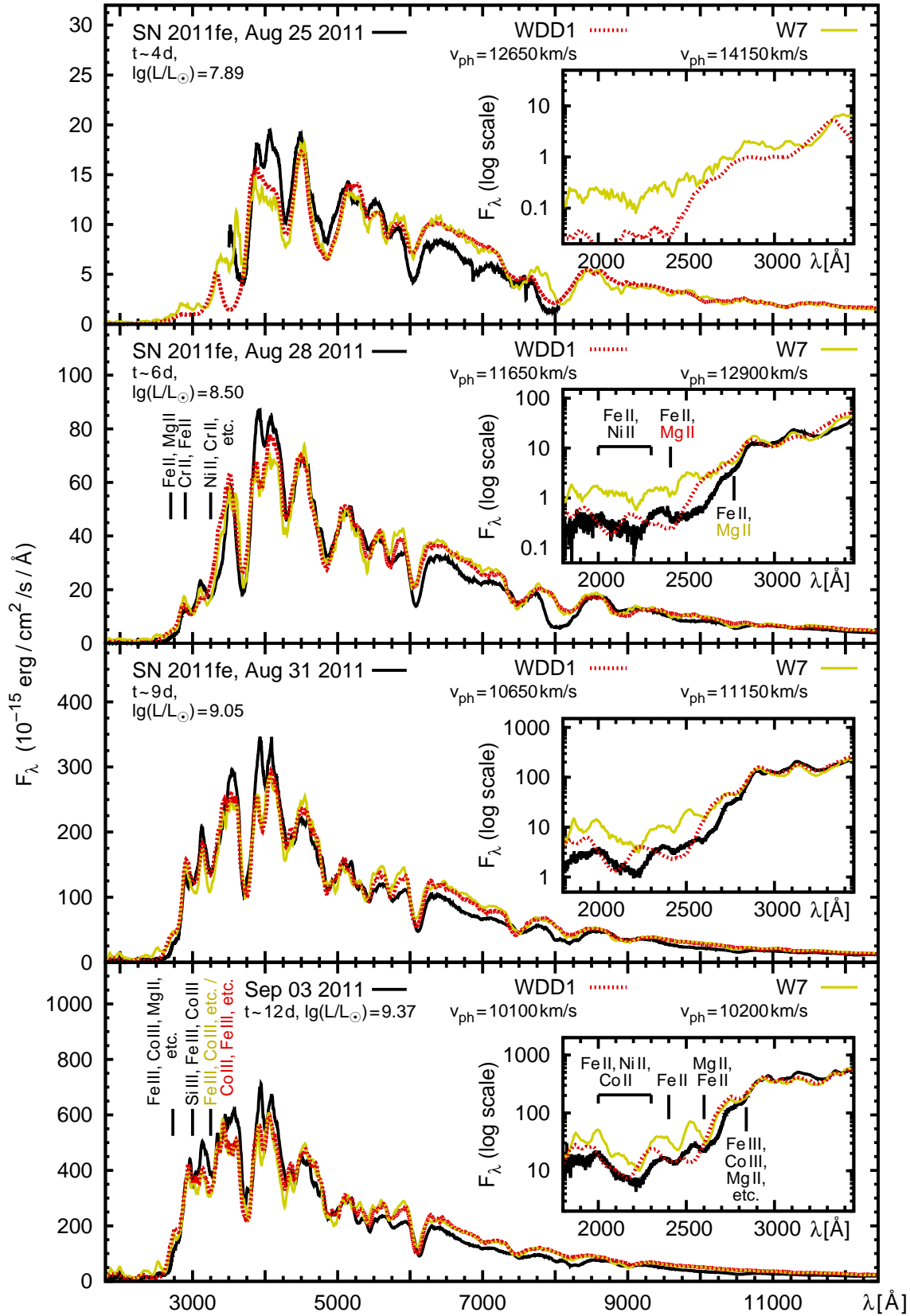}      
   \caption{Spectra of SN\,2011fe (black lines) compared to models
     based on the W7 and WDD1 density profiles (yellow\,/\,light grey,
     solid lines; red\,/\,grey, dotted lines, respectively). Insets
     show the UV in more detail (with logarithmic flux axis). The
     models have been reddened by
     $E(B-V)_\textrm{host}=\textrm{0.014}$\,mag. The ions responsible
     for the most prominent UV features in our model are marked, in
     order of importance. }
   \label{fig:sequence-11fe-w7wdd1comp}
\end{figure*}

\begin{figure*}   
   \centering
   \includegraphics[width=15.2cm]{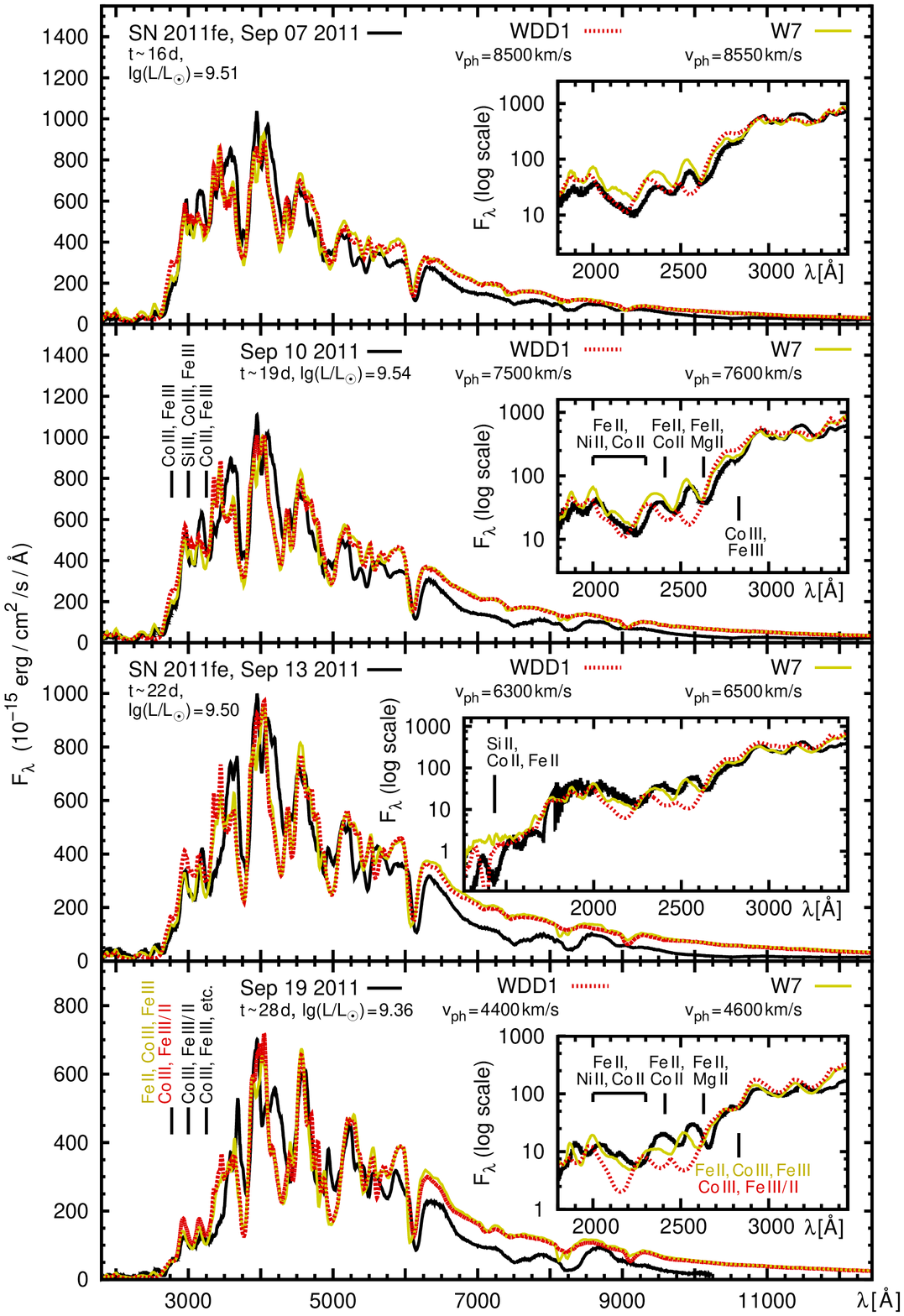}
   \caption{The W7 and WDD1 model sequences, continued.}
   \label{fig:sequence-11fe-w7wdd1comp-2}
\end{figure*}

Having used an early spectrum to determine \tr, we now analyse the \textit{HST}
time series using the W7 and WDD1 densities
(Figs~\ref{fig:sequence-11fe-w7wdd1comp} and
\ref{fig:sequence-11fe-w7wdd1comp-2}, respectively). Model parameters are
compiled in Table~\ref{tab:tvalues-11fe-standardmodels}. We do not use the last
three \textit{HST} spectra (Table \ref{tab:meanspec}), as our code is optimised
for the early photospheric phase, around and prior to maximum light, and these
spectra are later than 20 days after peak.

For the spectrum of 2011 August 25 we used $v_\textrm{ph}$ as used in the models
for the determination of \tr. For W7 we used an epoch $t=3.8$\,d and
$v_\textrm{ph} = 14150$\,\kms, for WDD1 $t=4.2$\,d and $v_\textrm{ph} =
12650$\,\kms. Despite the very early epoch and the high $v_\textrm{ph}$, strong
\SiII\ lines are visible, indicating the presence of burned material in the
outer layers. In order to reproduce these lines, the W7 models require $>30$ per
cent of the ejecta mass above $v_\textrm{ph}$ to be material heavier than
oxygen. This is only $\sim$13 per cent in the WDD1-based model.  Most of the
burned material consists of IMEs ($\textrm{9} \leq Z \leq \textrm{20}$), but
traces of Fe-group elements are also present. Line blanketing in the UV, which
is mostly due to overlapping lines of Fe-group elements, is much stronger in 
WDD1 than in W7, as the delayed-detonation places more material at high
velocity.  WDD1 yields a better reproduction of the peak around 4000\,\AA,
although neither model matches the feature well. This may be a consequence of
our assumption of black-body emission at the photosphere. However, more
sophisticated models that do not make that assumption also have difficulties
reproducing this high peak, which is seen in several SNe Ia.  Overall, both
models reproduce the observed features reasonably well except for \SiII\
$\lambda$6355, which is always too weak.  Increasing the Si abundance above the
photosphere to correct this results in too much high-velocity Si absorption at
later epochs.  Hints of a HVF are visible in the \CaII\ IR triplet, as confirmed
at later epochs.  As discussed above, we do not try to match these HVFs,
although they may be the reason for the strength of the \SiII\ line in this
spectrum.

Note that $v_\textrm{ph}$ is larger in the W7 model than in the WDD1 one. This
may seem counter-intuitive, as W7 has less mass at high velocity. The reason for
selecting a higher $v_\textrm{ph}$ for W7 is that because of the smaller mass at
high velocity, the UV opacity is smaller and hence the UV flux higher than in
WDD1 if similar $v_\textrm{ph}$ are used. By selecting a higher $v_\textrm{ph}$
the temperature is reduced and the UV flux decreases. This is obviously not an
ideal solution, since the line velocities are then too high, but it shows that
W7 does not have the correct density distribution at high velocities.

\begin{table}
\caption{Luminosities, photospheric velocities and temperatures of the
photospheric black-body in the W7 and WDD1 models.}
\label{tab:tvalues-11fe-standardmodels}
\centering
\begin{tabular}{ccclcc}
$\!\!$ Date $\!\!$ & $\!\!$ Phase $\!\!$ & $\!\!$ $t$ $\!\!$ & \lbol
& $v_\textrm{ph}$ & $T_\textrm{ph}$ \\ 
 (UTC) & (days) & (days) & $\!\!\!\!$($\textrm{10}^9 \Lsun$) & (\kms) & (K) \\
\hline \multicolumn{6}{c}{W7} \\ \hline
Aug. 25  & $-$15.3 & $\phantom{\textrm 0}$3.8 &   0.08  &  14150 &  10800 \\
Aug. 28  & $-$13.1 & $\phantom{\textrm 0}$6.0 &   0.32  &  12900 &  12100 \\
Aug. 31  & $-$10.1 & $\phantom{\textrm 0}$9.1 &   1.1	&  11150 &  14600 \\
Sept. 3  & $\phantom{\textrm 0}$$-$6.9 & 12.2 &   2.3   &  10200 &  15200 \\
Sept. 7  & $\phantom{\textrm 0}$$-$2.9 & 16.2 &   3.2   & 
$\phantom{\textrm{0}}$8550   & 15100 \\
Sept. 10 & $\phantom{\textrm 0}$$+$0.1 & 19.2 &   3.5   & 
$\phantom{\textrm{0}}$7600  & 14500 \\
Sept. 13 & $\phantom{\textrm 0}$$+$3.4 & 22.5 &   3.2   & 
$\phantom{\textrm{0}}$6500   & 14000 \\
Sept. 19 & $\phantom{\textrm 0}$$+$9.3 & 28.4 &   2.3   &  $\phantom{\textrm{0}}$4600
  & 13500 \\
\hline \multicolumn{6}{c}{WDD1} \\ \hline
Aug. 25  & $-$15.3 & $\phantom{\textrm 0}$4.2 &   0.08  &  12650 & 10300 \\
Aug. 28  & $-$13.1 & $\phantom{\textrm 0}$6.4 &   0.32  &  11650 & 12400 \\
Aug. 31  & $-$10.1 & $\phantom{\textrm 0}$9.5 &   1.1	&  10650 & 14500 \\
Sept. 3  & $\phantom{\textrm 0}$$-$6.9 & 12.6 &   2.3   &  10100 & 14800 \\
Sept. 7  & $\phantom{\textrm 0}$$-$2.9 & 16.6 &   3.2   & 
$\phantom{\textrm{0}}$8500   & 15100 \\
Sept. 10 & $\phantom{\textrm 0}$$+$0.1 & 19.6 &   3.5   &  $\phantom{\textrm{0}}$7500
  & 14800 \\
Sept. 13 & $\phantom{\textrm 0}$$+$3.4 & 22.9 &   3.2   & 
$\phantom{\textrm{0}}$6300   & 14400 \\
Sept. 19 & $\phantom{\textrm 0}$$+$9.3 & 28.8 &   2.3   &  $\phantom{\textrm{0}}$4400
  & 13600 \\
\hline
\end{tabular}
\end{table}

On 2011 August 28 the first UV spectrum was observed by \textit{HST}. Having the
UV available constrains our models significantly.  Although both density
profiles provide a good fit in the optical, the synthetic spectra show large
differences in the UV. In particular, the low flux level at $\lambda
\lesssim$\,2500\,\AA\ can only be reproduced if there is sufficient material at
high velocity to absorb photons, which is a feature of delayed-detonation models
\citep{iwa99}.  Features in the UV are dominated by singly-ionised Fe-group
elements and Mg (Fig.~\ref{fig:sequence-11fe-w7wdd1comp}). Ni, Fe,
Ti\,/\,V\,/\,Cr\footnote{The lines of Ti, V, and Cr are too heavily blended for
us to be able to distinguish the influence of each element. Therefore we
determined a combined abundance, with a mix similar to the results of
nucleosynthesis models \citep{iwa99}.} and Mg influence different parts of the
spectrum, so that their abundances can be determined independently. The
different UV features are discussed in more detail in
Section~\ref{sec:tomography-11fe-w7+}.

The next two spectra were obtained on 2011 August 31 and 2011 September 3
($\sim$\,9--13\,d after explosion) when the SN was significantly more luminous.
Between 2700 and 3500\,\AA, absorption by doubly-ionised Fe-group species now
appears. The photosphere is now in the incompletely burned region (10100 $-$
11150\,\kms), where the two models differ less in density. At both epochs, the
synthetic spectra match the observed ones reasonably well. Most differences are
in the UV.  The positions of the UV features of the W7 model agree with the
observed ones, but the overall flux is generally too high, implying too little
absorbing material, as discussed above.  The WDD1-based model reproduces the
overall flux, but the Fe-group lines are too blueshifted because of the
increased high-velocity absorption introduced by even a weak delayed-detonation
model. In both models, the regions just above the photosphere contains up to 6
per cent Fe by mass (this is so high that it is likely to be Fe synthesised in
the explosion), while the mass fraction of \Nifs, including the decay products
\Cofs\ and \Fefs\  (which is however negligible at this early phase), is between
11 and 30 per cent. The dominant constituents in this region are, however, IMEs.
The carbon abundance at these velocities has dropped to zero. A weak \CII\ line
may still be present at $\sim 6580$\,\AA\ in the observations: this can be
reproduced by the carbon in the outer layers.

By maximum light (the 2011 September 7 and 2011 September 10 spectra;
Fig.~\ref{fig:sequence-11fe-w7wdd1comp-2}), the photosphere has receded about
half way inside the ejecta ($v_\textrm{ph} \sim 7500-8600$\,\kms). The value of
\lbol\ at maximum is in good agreement with that derived by \citet{pereira13}.
In these layers \Nifs\ dominates, and little incompletely burned or unburned
material remains. In the UV, where the fit is very good, Co lines play a larger
role as \Nifs\ decays. The slight mismatch between models and observations in
the red and IR, where the models show a small flux excess, is probably a
consequence of the black-body approximation at the photosphere in a low electron
density environment (Section~\ref{sec:mont-radi-transf}).

At post-maximum epochs, (2011 September 13 and 19, $\sim$\,23 and 29\,d after
explosion, respectively, Fig.~\ref{fig:sequence-11fe-w7wdd1comp-2}), the largest
differences between W7 and WDD1 are again in the UV.  The W7  model now gives a
somewhat better match. In the near-UV, the absorptions of the WDD1 model are too
blue and the UV flux too low, again indicating too much opacity at high
velocities. The WDD1 model is in better agreement with the data in the far-UV
($\lambda < 2000$\,\AA) in the 2011 September 13 spectrum, the only epoch for
which far-UV data are available. However, it may be affected by too much
back-warming, again a consequence of the large mass at high velocities: this can
be seen in the excessive \SiIII\ 5740\,\AA\ absorption.

The September 19 spectrum can be modelled using $v_\textrm{ph} = 4600$\,\kms\
for W7 or 4400\,\kms\ for WDD1. The material at this depth is very rich in
\Nifs, and a more reliable analysis of these zones should be made using nebular
spectra.

\subsection{The `$\rho$-11fe' model}
\label{sec:tomography-11fe-w7+}

\begin{figure*}   
   \centering
   \includegraphics[width=15.0cm,angle=0]{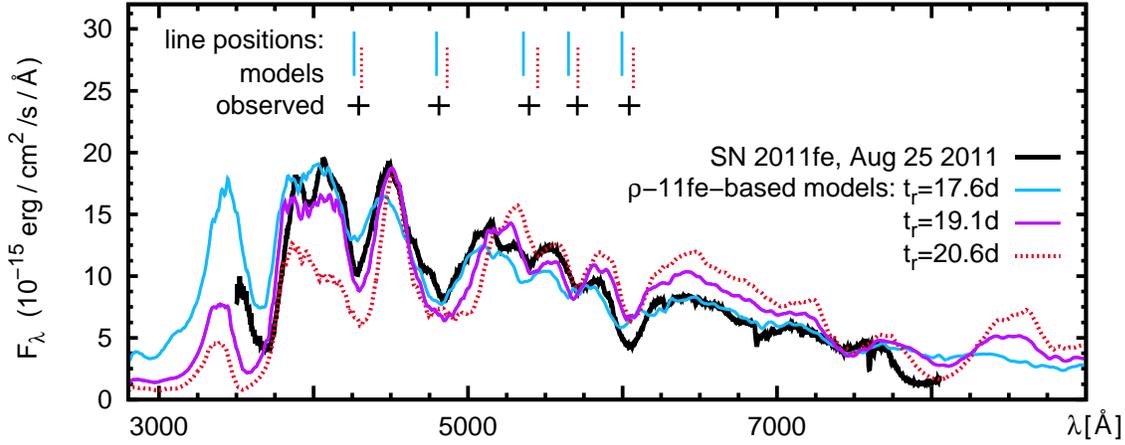}
   \caption{Early-time models for SN\,2011fe based on the $\rho$-11fe
     density profile, assuming different rise times \tr\ between
     17.6\,d and 20.6\,d. The optimum rise time is determined
     comparing the positions of prominent features in the optical
     between observed spectrum (black crosses) and models (light blue/grey, 
     solid lines for 17.6\,d; purple/dark grey, solid lines for 19.1\,d; 
     red/dark grey, dotted lines for 20.6\,d) 
     -- analogous to Fig.~\ref{fig:models-11fe-rt-iwamoto}.}
   \label{fig:models-11fe-rt-w7+}
\end{figure*}

\begin{table*}
  \caption{As Table~\ref{tab:linevel-measurements-iwamoto}, but for the
$\rho$-11fe model.}
\label{tab:linevel-measurements-w7+}
\centering
\begin{tabular}{lrrrrrr}
$t_r$ & 
$\ \ $ $\Delta\lambda$(Fe\,/\,Mg $\sim$\,4300\,\AA) $\!\!\!\!\!\!$ & 
$\ \ $ $\Delta\lambda$(Fe\,/\,etc. $\sim$\,4800\,\AA) $\!\!\!\!\!\!$ &
$\ $ $\Delta\lambda$(\SII\ $\lambda$5640) $\!\!$ &
$\!\!$ $\Delta\lambda$(\SII\ $\lambda$5972) $\!\!$ &
$\!\!$ $\Delta\lambda$(\SiII\ $\lambda$6355) $\!\!$ & 
$\!\!$ $\overline{\Delta\lambda}_\textrm{rel}$ $\!\!\!\!$\\ 
     & $\!\!\!\!$ (\AA) $\qquad$  &  $\!\!\!\!$ (\AA)
$\qquad$  & $\!\!\!\!$ (\AA) $\qquad$  & $\!\!\!\!$ (\AA) $\qquad$  & $\!\!\!\!$
(\AA) $\qquad$  & $\!\!\!\!$ (\%) $\!\!\!\!$  \\
\hline
\multicolumn{7}{c}{$\rho$-11fe-based models}\\
\hline
$\!\!\!\!$	17.6	$\qquad$ & $\!\!\!\!$	-29.0	$\qquad$ & $\!\!\!\!$	-15.8	$\qquad$ & $\!\!\!\!$	-37.9	$\qquad$ & $\!\!\!\!$	-57.2	$\qquad$ & $\!\!\!\!$	-47.8	$\qquad$ & $\!\!\!\!$	-0.70	$\!\!\!\!$	\\
$\!\!\!\!$	18.1	$\qquad$ & $\!\!\!\!$	-15.2	$\qquad$ & $\!\!\!\!$	-4.7	$\qquad$ & $\!\!\!\!$	-21.4	$\qquad$ & $\!\!\!\!$	-34.5	$\qquad$ & $\!\!\!\!$	-28.9	$\qquad$ & $\!\!\!\!$	-0.39	$\!\!\!\!$	\\
$\!\!\!\!$	18.6	$\qquad$ & $\!\!\!\!$	0.1	$\qquad$ & $\!\!\!\!$	3.7	$\qquad$ & $\!\!\!\!$	1.5	$\qquad$ & $\!\!\!\!$	-27.5	$\qquad$ & $\!\!\!\!$	-14.5	$\qquad$ & $\!\!\!\!$	-0.12	$\!\!\!\!$	\\
$\!\!\!\!$	19.1	$\qquad$ & $\!\!\!\!$	12.0	$\qquad$ & $\!\!\!\!$	15.9	$\qquad$ & $\!\!\!\!$	13.6	$\qquad$ & $\!\!\!\!$	-20.0	$\qquad$ & $\!\!\!\!$	10.9	$\qquad$ & $\!\!\!\!$	0.14	$\!\!\!\!$	\\
$\!\!\!\!$	19.6	$\qquad$ & $\!\!\!\!$	21.2	$\qquad$ & $\!\!\!\!$	21.2	$\qquad$ & $\!\!\!\!$	30.7	$\qquad$ & $\!\!\!\!$	-11.9	$\qquad$ & $\!\!\!\!$	7.8	$\qquad$ & $\!\!\!\!$	0.28	$\!\!\!\!$	\\
$\!\!\!\!$	20.1	$\qquad$ & $\!\!\!\!$	23.8	$\qquad$ & $\!\!\!\!$	41.2	$\qquad$ & $\!\!\!\!$	44.4	$\qquad$ & $\!\!\!\!$	-4.4	$\qquad$ & $\!\!\!\!$	18.5	$\qquad$ & $\!\!\!\!$	0.49	$\!\!\!\!$	\\
$\!\!\!\!$	20.6	$\qquad$ & $\!\!\!\!$	18.7	$\qquad$ & $\!\!\!\!$	52.5	$\qquad$ & $\!\!\!\!$	55.6	$\qquad$ & $\!\!\!\!$	3.7	$\qquad$ & $\!\!\!\!$	26.5	$\qquad$ & $\!\!\!\!$	0.61	$\!\!\!\!$	\\
\hline
\multicolumn{7}{c}{$\!\!\!\!$regression line fit to
$\overline{\Delta\lambda}_\textrm{rel}$--\tr\
relation $\Rightarrow \overline{\Delta\lambda}_\textrm{rel}\simeq\textrm{0}$ for
$t_{r,\rho\textrm{-11fe}}\simeq\textrm{19.0}$\,d$\!\!\!\!$}\\
\hline
\end{tabular}
\begin{flushleft}
\end{flushleft}
\end{table*}

Spectra produced with the W7 density profile have too much flux in the UV. On
the other hand, spectra produced with WDD1 have the correct UV flux level but
the UV features are typically too blue when compared to the observations. It is
interesting however to notice that the two models are practically
indistinguishable in the optical \citep[see also][]{roepke12}, highlighting the
importance of the UV as a diagnostic. Previous modelling of combined UV/optical
spectra of SNe\,Ia \citep[\eg][]{sau08, wal12} suggests that the discrepancies
are due to the lack of material at high velocity in the case of W7, and to an
excess of such material in the case of WDD1. In fact, the layer with $v >
24000$\,\kms\ is the region where the two models disagree most strongly (see
Fig. 3). We therefore constructed arbitrary density structures that try to
address these issues, and tested their spectral behaviour. We show here the
results of the best of such models, which we call ``$\rho$-11fe''. This is
essentially a rescaled version of WDD1 with a steeper high-velocity tail and
therefore less mass at the highest velocities. Thus, $\rho$-11fe may be thought
of as a very weak delayed detonation, although other models may show similar
properties. The model is shown in Fig.3 along with W7 and WDD1. Formally,
$\rho$-11fe has $\KE = 1.23\times10^{51}$\,erg, which is less than both W7 and
WDD1, but here we can only test the outer part of the ejecta, so this result
must be taken with caution. The high-velocity layers of $\rho$-11fe carry little
mass (0.003\,\Msun\ above 24000\,\kms) and energy ($0.02\times10^{51}$\,erg in
the same zone). 


As a first check of $\rho$-11fe, we performed the same rise-time test as for W7
and WDD1 (Section~\ref{sec:models-risetime-iwamoto}), setting $T_\textrm{ph} =
10800$\,K, which seems to give an optimal match to the 25 Aug spectrum.
Synthetic spectra for the different rise times are shown in
Fig.~\ref{fig:models-11fe-rt-w7+} and line offsets are given in Table
\ref{tab:linevel-measurements-w7+}.  We allowed generously large timesteps
between the various models.  While the longer risetime clearly results in a very
red spectral distribution, incompatible with the observations, the two shorter
risetimes yield reasonable results. UV data would increase the leverage of this
technique. If we test the position of the absorption features we infer an
optimum rise time of $\tr=19.0\pm0.5$\,d, similar to the values obtained
for W7 and WDD1. We modelled the spectral time series using this value.

\begin{figure*}   
   \centering
   \includegraphics[width=15.2cm]{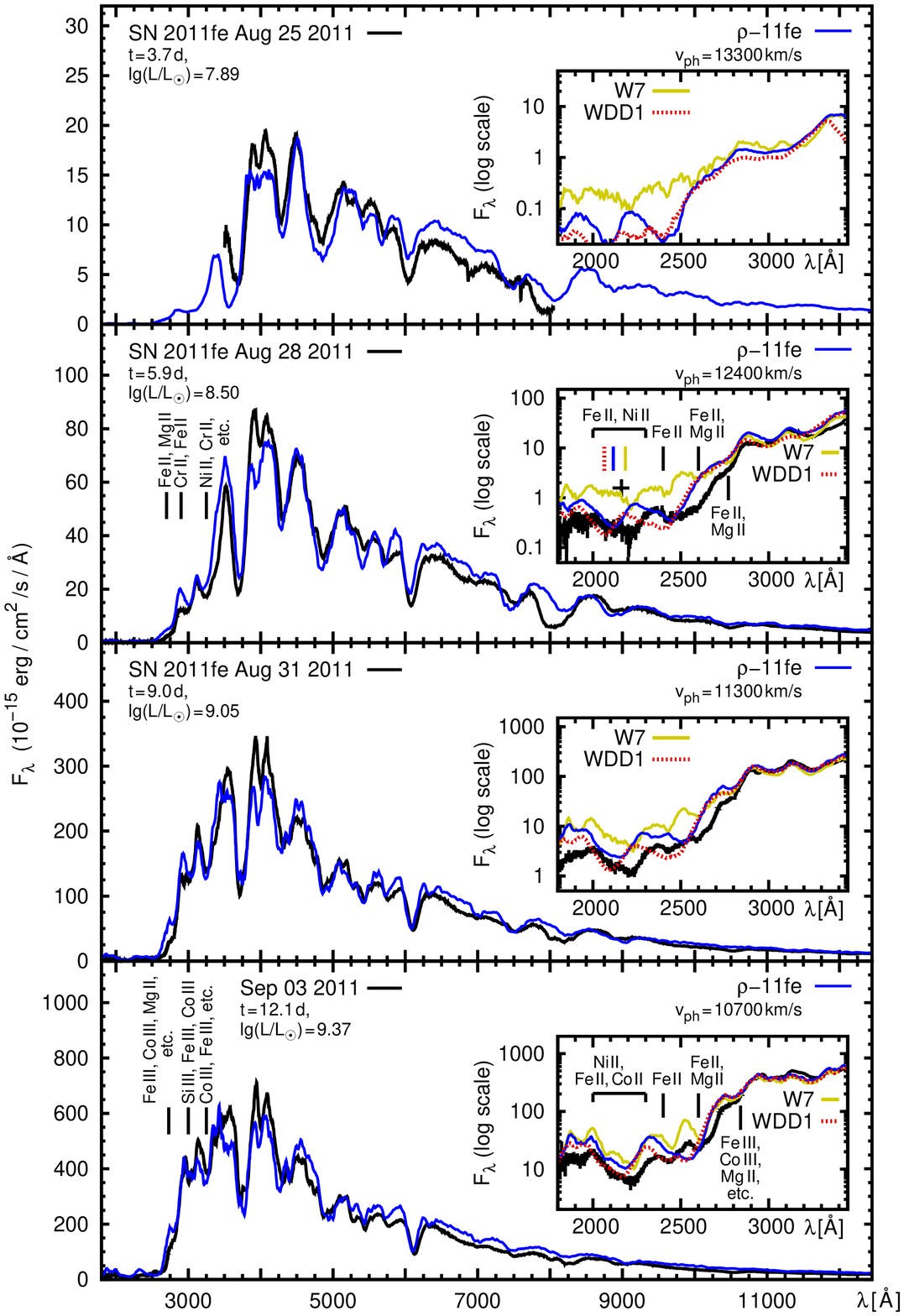}   
   \caption{Spectra of SN\,2011fe (black lines) compared to models
     based on the $\rho$-11fe density profile (blue, solid lines). The
     insets also show the W7-based models (yellow, solid lines) and the 
     WDD1-based models (red, dotted lines) for comparison. The inset for Aug 28 
     demonstrates that $\rho$-11fe reproduces the observed line positions 
     (observed centroid of 2000--2300\,\AA\ feature marked with a cross;
     model feature centroids marked with tics) somewhat better than WDD1. W7
     reproduces the line positions practically perfectly, but has too low a UV
     opacity to reproduce the observed UV flux.}
   \label{fig:sequence-11fe-w7plus}
\end{figure*}

\begin{figure*}   
   \centering
   \includegraphics[width=15.2cm]{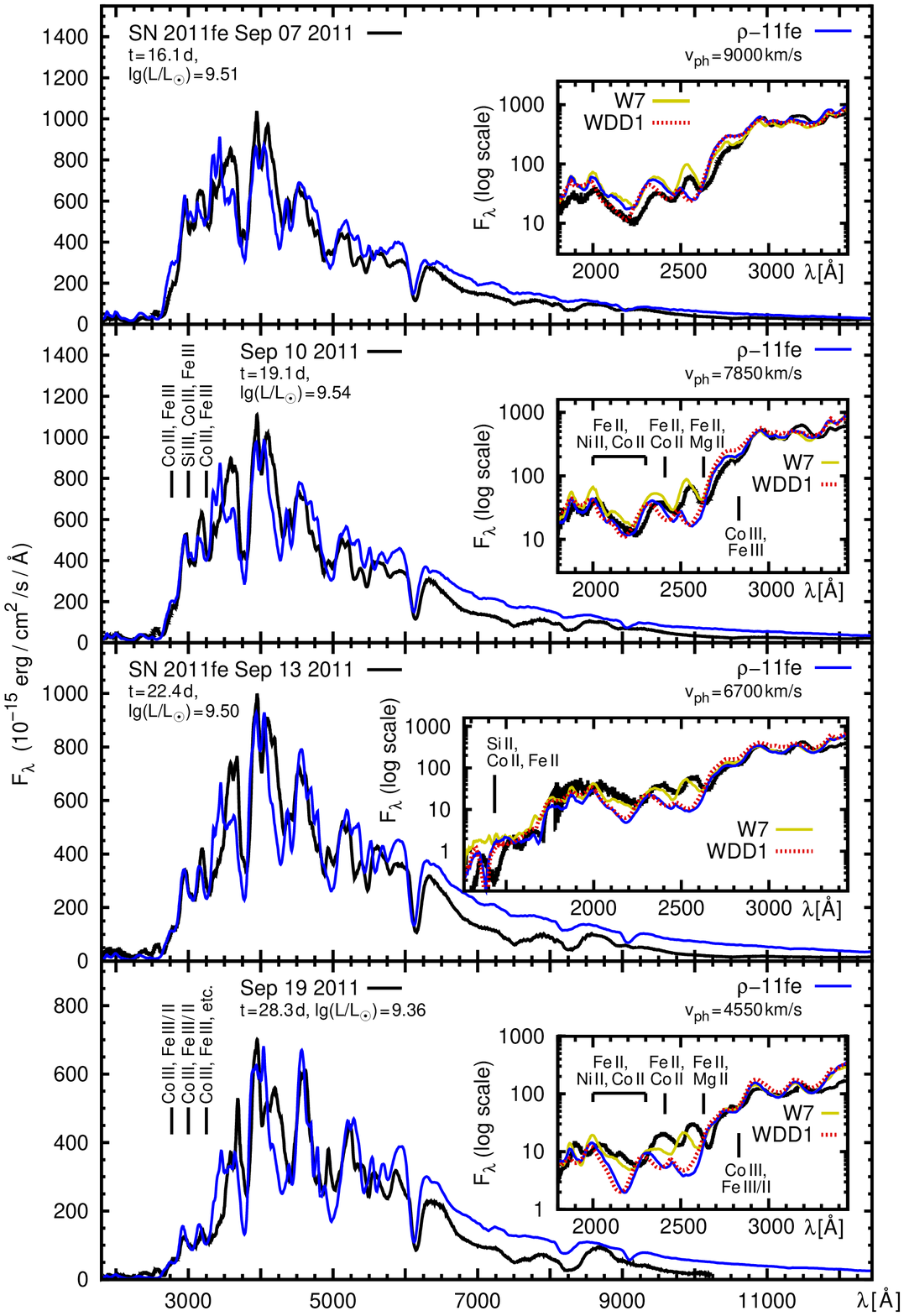}
   \caption{The $\rho$-11fe model sequence, continued.}
   \label{fig:sequence-11fe-w7plus-2}
\end{figure*}

The synthetic spectra for the model series are shown in
Figs~\ref{fig:sequence-11fe-w7plus} and \ref{fig:sequence-11fe-w7plus-2}. Model
parameters and physical properties are listed in
Table~\ref{tab:tvalues-11fe-w7plus}. As expected, the $\rho$-11fe model spectra
are intermediate between the W7 and WDD1 spectra. They show an overall better
reproduction of the spectral distribution. The largest differences with respect
to the W7 and WDD1 models are in the UV, but it is interesting that several
features in the optical are also improved.

At the earliest epochs (2011 August 25, 28 and 31), $\rho$-11fe shows
stronger UV absorption than W7, but the features are formed at somewhat longer
wavelength than with WDD1, improving the quality of the fit. Further
optimisation of the density profile may lead to an even better match. This would
require an accurate study of the density in the outermost layers, including the
modelling of HVFs, which we defer to later work. Additionally, 3D effects may
not be negligible in the outermost layers \citep{tanaka06}. 

Since the ad-hoc density profile we constructed yields improved fits to the
UV-optical spectra up to several days after maximum, we can use it to diagnose
abundances in the outer layers of SN\,2011fe. 
The Ni/Co/Fe blend near 2000--2300\,\AA\ is a diagnostic for radioactive
material from the \Nifs\,$\rightarrow$\,\Cofs\,$\rightarrow$\Fefs\ decay chain,
and its shape shows some sensitivity to \Nifs\ even in the outer ejecta.  The Fe
abundance can be constrained from Fe-dominated features such as the near-UV
blend near 3300\,\AA\ and from the optical Fe features. If the Fe abundance
turns out to be larger than the estimated abundance in the progenitor, the
abundance of stable Fe synthesised directly in the explosion can also be tested,
since at early times little \Nifs\ has decayed to \Fefs. The UV spectra thus
allow us to constrain the abundances of both Fe and \Nifs\ in the outer part of
the SN ejecta \citep[cf.][]{hachinger13}.

\begin{table}
  \caption{Luminosities, photospheric velocities and temperatures of the
    photospheric black-body in the $\rho$-11fe model (analogous to Table
    \ref{tab:tvalues-11fe-standardmodels}).}
\label{tab:tvalues-11fe-w7plus}
\centering
\begin{tabular}{ccclcc}
$\!\!$ Date $\!\!$ & $\!\!$ Phase $\!\!$ & $\!\!$ $t$ $\!\!$ &  \lbol & 
$v_\textrm{ph}$ & $T_\textrm{ph}$ \\ 
 (UTC) & (days) & (days)  & $\!\!\!\!$($\textrm{10}^9 \Lsun$) & (\kms) & (K) \\
\hline \multicolumn{6}{c}{$\rho$-11fe} \\ \hline
Aug. 25  & $-$15.3 & $\phantom{\textrm 0}$3.7 &   0.08  & 13300 &  10800 \\
Aug. 28  & $-$13.1 & $\phantom{\textrm 0}$5.9 &   0.32  & 12400 &  12100 \\
Aug. 31  & $-$10.1 & $\phantom{\textrm 0}$9.0 &   1.1   & 11300 &  14500 \\
Sept. 3  & $\phantom{\textrm 0}$$-$6.9 & 12.1 &   2.3   & 10700 &  14900 \\
Sept. 7  & $\phantom{\textrm 0}$$-$2.9 & 16.1 &   3.2   & 
$\phantom{\textrm{0}}$9000 & 15100 \\
Sept. 10 & $\phantom{\textrm 0}$$+$0.1 & 19.1 &   3.5   &  
$\phantom{\textrm{0}}$7850 & 14700 \\
Sept. 13 & $\phantom{\textrm 0}$$+$3.4 & 22.4 &   3.2   & 
$\phantom{\textrm{0}}$6700 & 14100 \\
Sept. 19 & $\phantom{\textrm 0}$$+$9.3 & 28.3 &   2.3   &  
$\phantom{\textrm{0}}$4550 & 13500 \\
\hline
\end{tabular}
\end{table}

\subsection{Abundances}
\label{sec:tomography-11fe-w7+-abundances}

\begin{figure*}   
  \centering
  \includegraphics[angle=270,width=13.5cm]{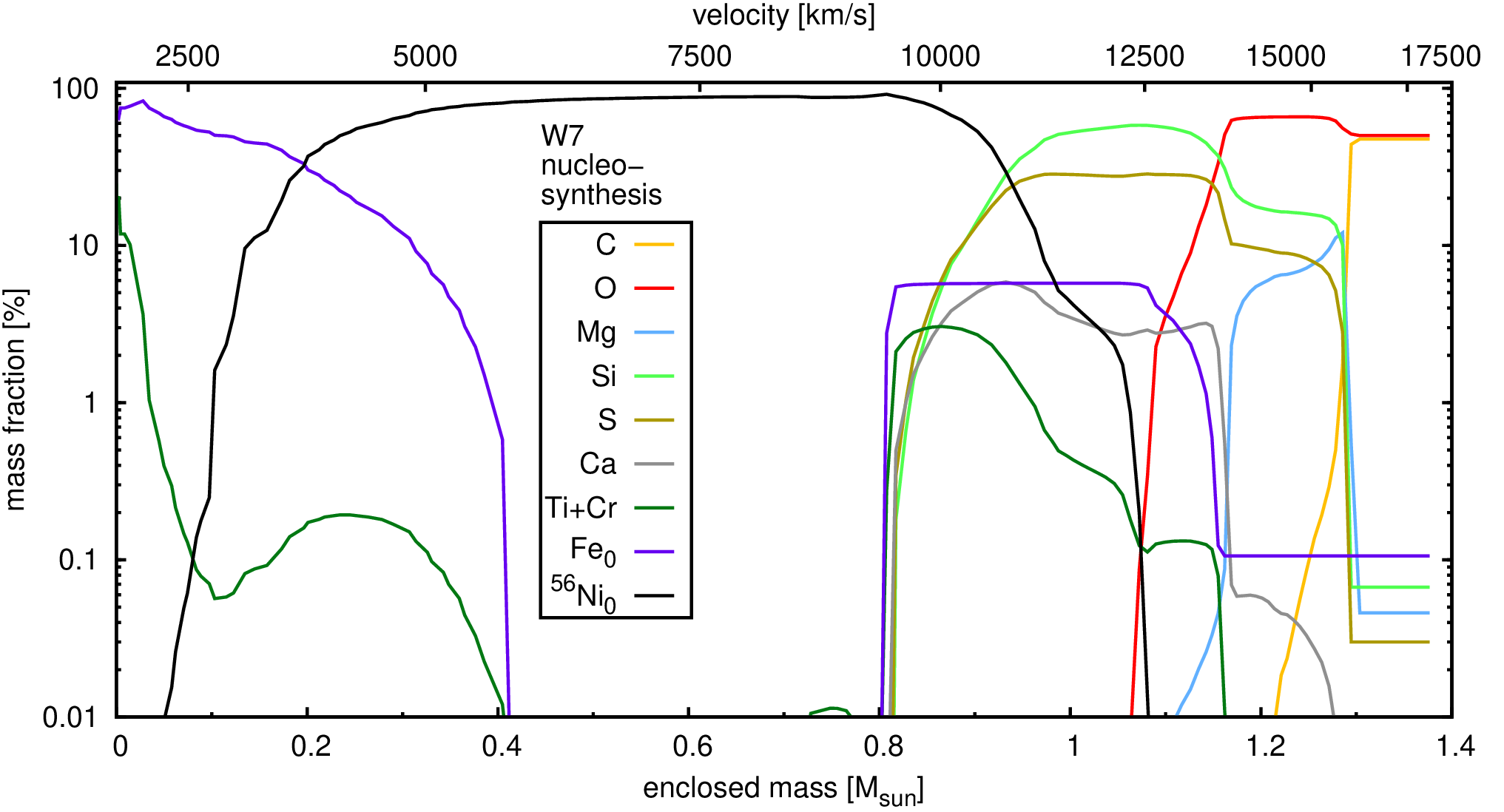}
  \\[0.15cm]
  \includegraphics[angle=270,width=13.5cm]{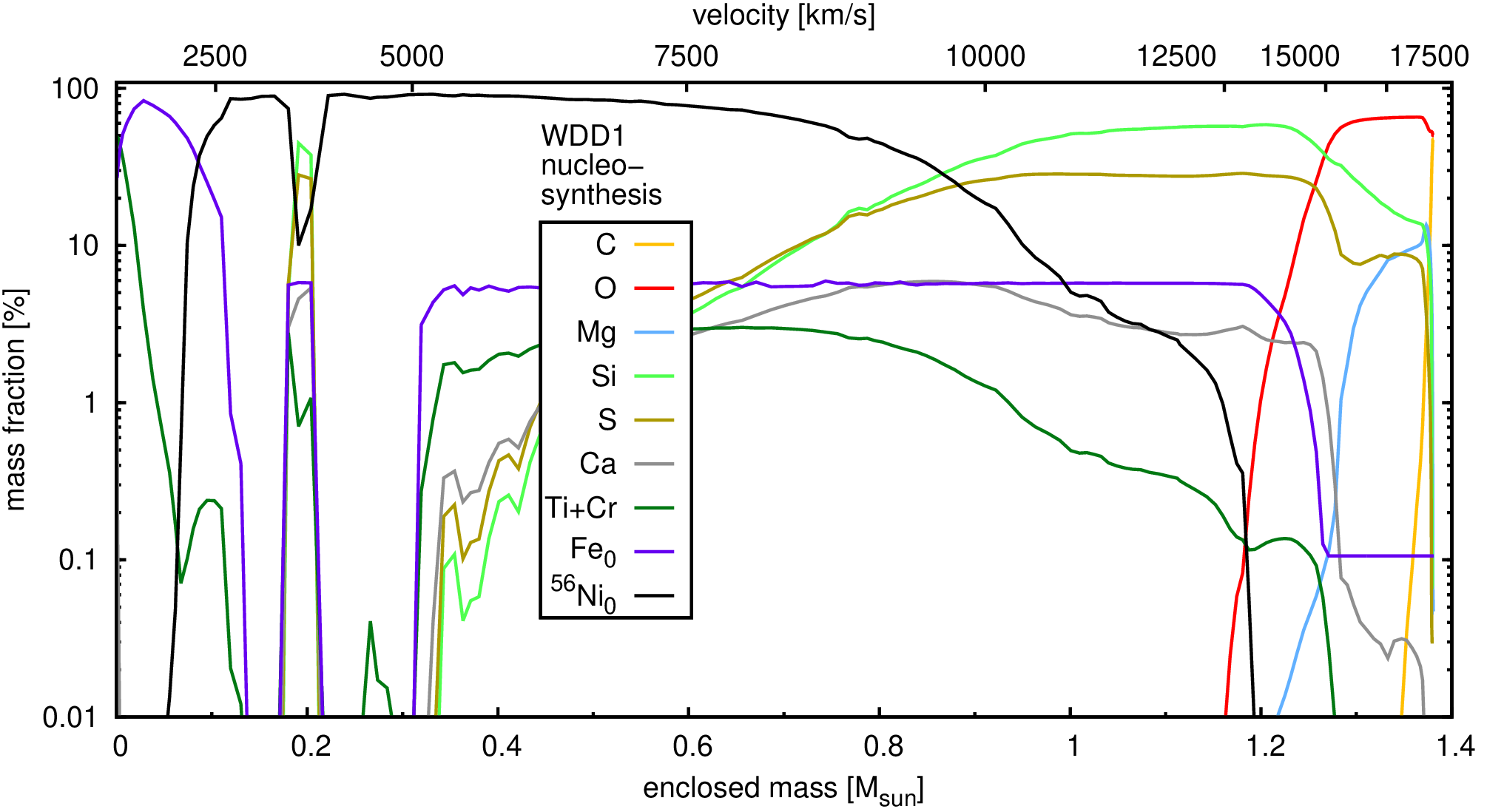}
  \\[0.15cm]
  \includegraphics[angle=270,width=13.5cm]{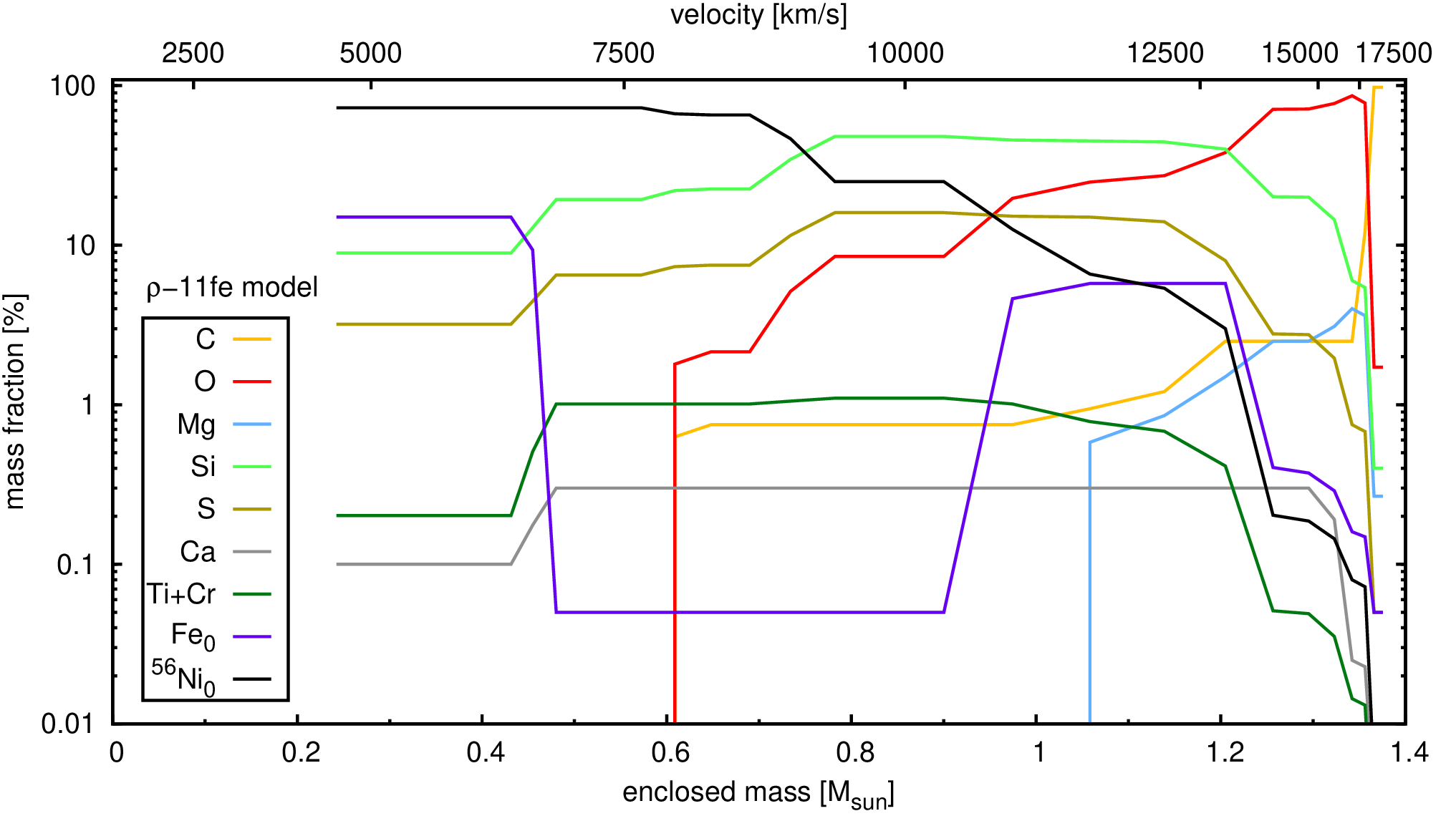}
  \caption{Abundances of W7 and WDD1 nucleosynthesis calculations
    \citep[][\textit{top and middle panel}, plotted in mass space]{iwa99},
    compared to our tomography based on $\rho$-11fe (\textit{lower panel}).
    The Ni/Co/Fe abundances are given in terms of the mass fractions
    of \Nifs\ and stable Fe at $t=0$ [$X(^{56}\textrm{Ni}_0)$,
    $X(\textrm{Fe}_0)$]; stable Ni and Co are present only in traces
    (few per cent or less by mass) in our model.}
  \label{fig:abundances-11fe}
\end{figure*}

The one-dimensional abundance stratification we infer for SN\,2011fe using the
$\rho$-11fe profile is shown in Fig.~\ref{fig:abundances-11fe} (bottom panel).
Although it bears similarities to the nucleosynthesis of W7, a 1D deflagration 
model (\citealt{iwa99}; Fig.~\ref{fig:abundances-11fe}, top panel), there are
significant differences in the degree of mixing, in particular at intermediate
velocities, the composition of which is very reminiscent of WDD1, a 1D
delayed-detonation model (\citealt{iwa99}; Fig.~\ref{fig:abundances-11fe},
middle panel). 3D delayed-detonation models currently do not seem to agree with
similar models in 1D \citep{seit13}.

A thin outer layer above $\sim$19400\,\kms\ introduced as an extra abundance zone
above the August 25 photosphere, containing $\sim$0.01\,\Msun, is composed almost
exclusively of carbon (98 per cent by mass). A higher oxygen abundance in these
layers would lead to spurious absorption in the blue wing of the \OI\
$\lambda$7773 feature, as discussed in Section~\ref{sec:carbon-layer}. We tested
the sensitivity of the models to the Fe abundance in the outermost layers, and
found that a sub-solar abundance gives the best results (see Sect. 6.1). This is
consistent with the metallicity of M101 \citep[][]{stoll11}.

The layer immediately below ($16000<v\leq19400$\,\kms) contains predominantly
oxygen (87 per cent by mass). The carbon abundance is low there (2.5 per cent). 
Some IMEs (e.g. Si, 6 per cent) and even traces of \Nifs\  (0.1 per cent
including decay products \Cofs\ and \Fefs) are also present, as is
directly-synthesised Fe (0.2 per cent) and other Fe-group elements. These are
required in order to reproduce the observed lines in the August 25 spectrum. 

The layer at $13300 < v \leq 16000$\,\kms\ contains a higher fraction
of burning products, including Si (20 per cent by mass),
directly-synthesised Fe (0.4 per cent), and \Nifs\ (0.2 per cent).

IMEs dominate the composition at intermediate velocities, between 9000 and
13300\,\kms. The most abundant IME is Si, as expected. Below 9000\,\kms, IMEs
decrease, and \Nifs\ becomes the dominant species. Its abundance reaches $\sim
65$ per cent below 9000\,\kms\ and it increases further at lower velocities. 
The ratio of \Nifs\ to stable Fe above 11000\,\kms\ appears to be somewhat
higher than in W7 or WDD1. This may be regarded as an indication of a sub-solar
metallicity of the progenitor \citep{iwa99}. Some carbon and oxygen may be
present down to 8000\,\kms, with very low abundance \citep{parrent12}.  The
upper limit is set by the strength of the line near 6600\,\AA\ and by the
appearance of the feature near 7200\,\AA\ \citep{mazzali2001}.

The photosphere of the last spectrum analysed here is at 4550\,\kms. Below this
velocity a core of $\sim 0.24$\,\Msun\ should exist, which can be analysed by
modelling the nebular spectra \citep[e.g.][]{ste05}. Above this velocity we find
$\sim 0.4$\,\Msun\ of \Nifs\ in a total mass of $\sim 1.14$\,\Msun.  As we
cannot establish the exact composition of the core or its mass based on the
analysis of photospheric-phase spectra only  
\citep[see, \eg,][]{mazzali11,mazzali12}, we can only state that our model may
contain 0.4\,$-$\,0.7\,\Msun\ of \Nifs.  This is in good agreement with the W7
and WDD1 models, and with the \Nifs\ mass estimate from the bolometric
luminosity of SN\,2011fe \citep[$0.53\pm0.11$\,\Msun;][]{pereira13}. It would
not be at all surprising, in fact, if the inner part of the ejecta were
dominated by stable Fe-group elements \citep{hoeflich04,maz07}. On the other
hand, an accurate estimate of the properties of the inner ejecta must await
appropriate modelling of the nebular spectra.

\section{Discussion}
\label{sec:discussion}

\subsection{Fe-group abundances and progenitor metallicity}
\label{sec:fe-group-abundances}

The distribution of Fe-group elements in SNe Ia largely determines their light
curves and greatly affects their spectra. It is therefore a key element when
comparing theoretical models to observations
\citep[e.g.][]{maz00,maz01lc,maz07,roe07b}. Different SN Ia explosion models
(deflagrations, delayed detonations, double-degenerate mergers) with different
progenitor metallicities will differ in their average \Nifs\ and stable Fe-group
yield and/or in the distribution of these elements.  The Fe-group content of
SNe\,Ia with good temporal and spectral coverage can be estimated with high
precision \citep[e.g.][]{ste05,stritzinger06b,tanaka11,hachinger13}.

Earlier studies suggested that the UV is the spectral region from which the
abundance of \Nifs\ and other Fe-group elements in the outer layers of the SN
can best be determined \citep[e.g.][]{lentz00a,wal12,hachinger13}.  Very early
light-curve data may also help in this respect \citep{piro13b}. 
\citet{foley13a} derived a sub-solar abundance for the progenitor of SN\,2011fe
by evaluating models of \citet{lentz00a} and employing a semi-empirical argument
based on the metallicity dependence of the amount of \Nifs\ versus stable
Fe-group elements produced in the explosion. However, this method accounts
neither for the different \Nifs\ production in SNe Ia with different
luminosities at fixed progenitor metallicity, nor for the actual Fe-group
opacities in SN\,2011fe, which can differ from the opacities used by
\citet{lentz00a}. With our detailed models for SN\,2011fe we can directly
investigate the metal content in the outer ejecta, where nucleosynthesis is
expected to play a minor role and the progenitor metallicity may still be
reflected in the Fe-group content.

\begin{figure*}   
   \centering
   \includegraphics[width=15.2cm]{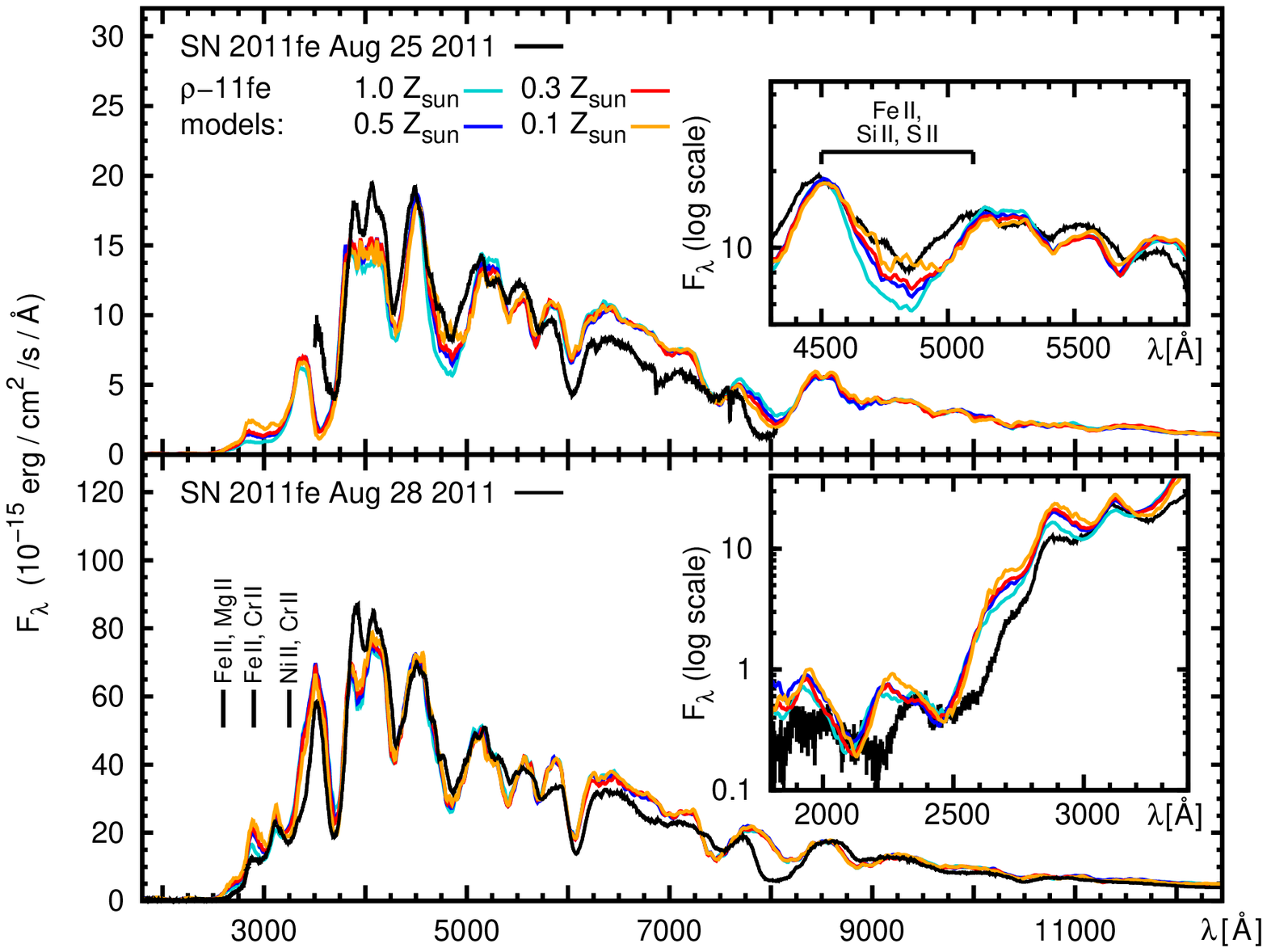}   
   \caption{Effect of varying the Fe-group abundances in the outermost layers:
     earliest spectra of SN\,2011fe compared to synthetic spectra from the 
     $\rho$-11fe model sequence computed for different Fe-group abundances in 
     the outermost layers ($v>\textrm{19400}$\,\kms). The insets show the
     UV in more detail for Aug 28 and the Fe-dominated blend
     around 4800\,\AA\ for Aug 25, respectively.}
   \label{fig:sequence-11fe-outerZ}
\end{figure*}

We tested the Fe-group content in the outermost layers by re-computing the
$\rho$-11fe-based spectral models and scaling the Fe-group abundances at
$v>\textrm{19400}$\,\kms\ to different fractions of the solar values
\citep{asplund09}. Fig.~\ref{fig:sequence-11fe-outerZ} shows that the effect of
metallicity is stronger in the UV but more linear in the optical, probably
because the latter is less saturated.

The Fe-dominated absorption at $\sim$4800\AA\ in the August 25 spectrum is too
strong in the model with solar metallicity (Fig.~\ref{fig:sequence-11fe-outerZ},
top panel, inset). This absorption is still present if $Z = 0.1$, but in this
case it is practically only caused by Fe located near the photosphere, which is
at 13300\,\kms.  Si and Mg also contribute significantly, and the shape of the
synthetic feature does not match that of the observed one. Actually, this is the
only portion of the optical spectrum that is as sensitive to metallicity as the
UV, or more.  A synthetic spectrum where the Fe-group abundances in the outer
layers are set to the metallicity of M101, $Z \simeq 0.5 Z_{\odot}$, matches the
UV-optical spectrum reasonably well. It shows less absorption in the optical Fe
feature, which makes it preferable over the $Z_{\odot}$ model, while the UV flux
level does not increase significantly.  The model with $Z = 0.3 Z_{\odot}$ is
similarly good.  All models show too much flux between 2500 and 3000\,\AA\ on
Aug 28. This probably reflects remaining shortcomings of $\rho$-11fe.

In order to test the sensitivity of our models to the assumed parameters, we
ran models where we changed the velocity of the photosphere by $\pm 400$\,\kms,
a range which represents a realistic uncertainty in this parameter.  The change
alters the temperature of the lower boundary black-body (deeper photospheres
are hotter), but the models yield very similar UV fluxes. This confirms that
the flux we see in the UV is not the flux at the lower boundary, but rather
radiation re-emitted by metals within the model atmosphere, as already stated
by \citet{maz00}. 


As a general remark, the sensitivity of the UV spectra to the (virtually)
unburned outer layer will crucially depend on that layer's extent.  This is
evident when comparing the present work with our earlier study of maximum-light
spectra \citep{wal12}. In those models a larger fraction of the ejecta was
considered unburned and tracked the progenitor metallicity. This resulted in a
stronger dependence of the spectra on metallicity. 

The UV spectra can also be used to diagnose the Fe-group composition in the
incompletely burned layers ($\sim$\,9000\,$-$\,13000\,\kms), where we find
mostly \Nifs\ and some directly-synthesised Fe 
(Section \ref{sec:tomography-11fe-w7+-abundances}).  This relative composition 
appears to lie between the solar-metallicity model W7 (where only \Feffo\ is
produced in partially-burned layers because of excess neutrons in $^{22}$Ne) and
the zero metallicity explosion model `W70' \citep{iwa99}, where only \Nifs\ is
produced. This indirectly suggests that the metallicity of the progenitor of
SN\,2011fe was moderately sub-solar.

\begin{figure*}   
   \centering
   \includegraphics[width=15.2cm]{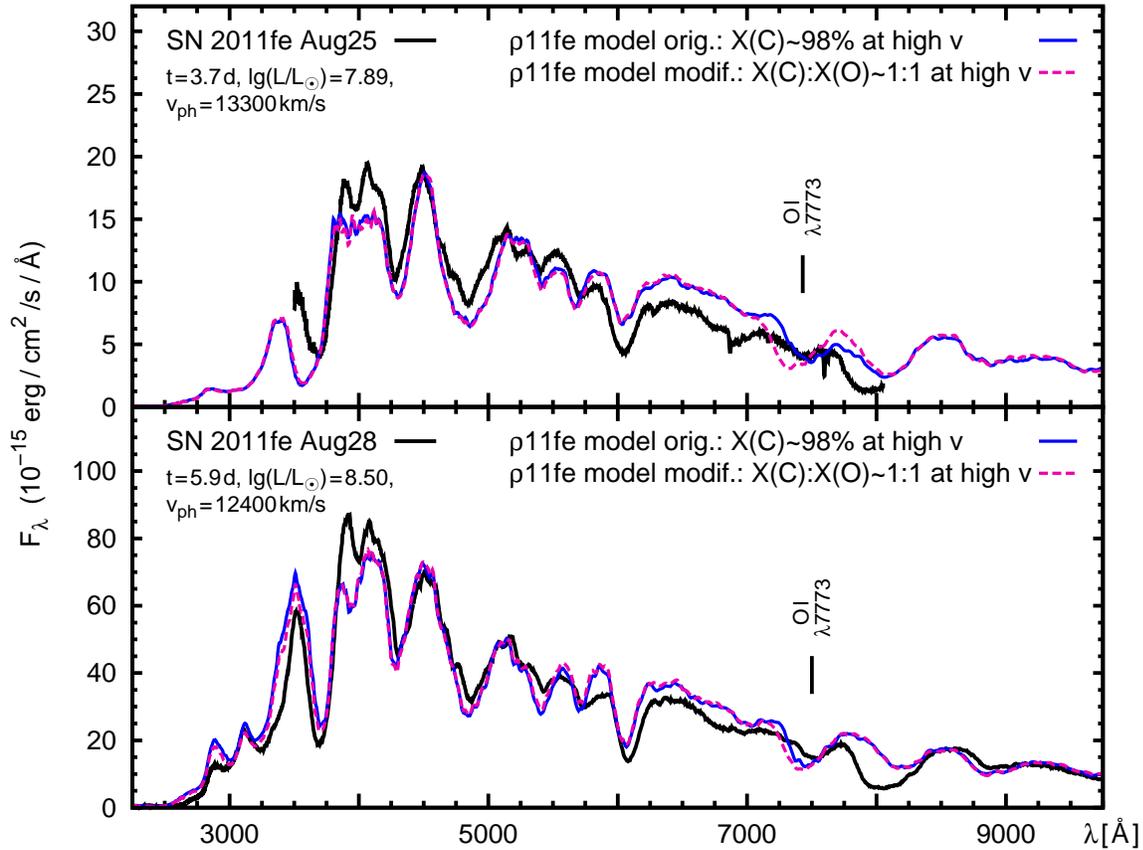}   
   \caption{Effect of oxygen in the outermost layers of the model: the
     earliest two spectra of SN\,2011fe (black) compared to $\rho$-11fe
     models. A spectrum where the outer layer is almost exclusively
     composed of carbon ($\sim$\,98 per cent by mass, blue\,/\,grey,
     solid lines) fits the observed \OI\ $\lambda$7773 line reasonably well, 
     much better than a model where C and O have the same abundances at
     $v>\textrm{19400}$\,\kms\ (red\,/\,grey, dotted lines).
     In the later spectrum the difference is already much smaller.}
   \label{fig:sequence-11fe-outercarbon}
\end{figure*}

\subsection{Estimate of rise-time and constraints on the progenitor system}
\label{sec:estim-t_r-constr}

Spectral modelling can be used to estimate the epoch of early spectra through
the combined fitting of temperature and line velocity. Our results indicate that
the rise-time is larger than estimated by simple analytical fits to the light
curve, regardless of the model used. This confirms that diffusion of radiation
delays the rise of the light curve. Such a delay was already derived for
SN\,2010jn \citep[1.0\,d;][]{hachinger13}. This was a very luminous SN Ia, in 
which \Nifs\ was closer to the surface \citep{hachinger13} than in SN 2011fe
which was less luminous and declined more rapidly \citep[see also][who showed
that SNe with a broader distribution of 56Ni as derived from the nebular
emission lines also have a broader light curve]{mazzali98, maz07}. The diffusion
time of the first photons should therefore be smaller, and the delay smaller. 
The slightly larger delay in SN\,2011fe (1.4\,d) is consistent with a smaller
\Nifs\ mass, and with \Nifs\ being located deeper in the ejecta.

The assumed rise time \tr\ has implications for the constraints on the
progenitor radius determined from the earliest light-curve points.
\citetalias{nugent11} used their first light-curve point (MJD 55797.2) and the
earlier non-detections (Section~\ref{sec:phot-prop}), together with the
analytical early-emission models of \citet{rabinak12} and \citet{kas10}, to
constrain the progenitor radius to $R_\textrm{prog} \lesssim
\textrm{0.1\,\Rsun}$.  Taking into account a later non-detection ($g>19.0$ at
MJD 55796.9), \citet[][hereafter B12]{bloom12} derived $R_\textrm{prog}\lesssim
0.02$\,\Rsun.

In our $\rho$-11fe model, the preferred explosion time is 1.4\,d earlier than
the date inferred by \citetalias{nugent11}, giving an explosion epoch of MJD
55795.3. This makes the limit of $g>22.2$ (cf.\ Section~\ref{sec:phot-prop}) at
MJD 55796.2 the most relevant data point for constraining the progenitor radius.
By MJD 55796.2, the SN has had about 22\,h for a possible post-shock cooling
lightcurve to fade \citep{piro13b}. Following the analysis of
\citetalias{bloom12} and using the models of \citet{rabinak11}, we derive
$R_\textrm{prog} \lesssim 0.02$\,\Rsun, a very similar constraint to
\citetalias{bloom12}. That is, a non-detection of $g>19.0$ at 4\,h after
explosion formally gives similar constraints as a non-detection of $g>22.2$ at
22\,h after explosion.

However, as discussed by \citet*{rabinak12} and \citetalias{bloom12}, the
expressions used for the early SN luminosity from \citet{rabinak11} assume that
the post-shock pressure is dominated by radiation.  When the shock diffusion
front reaches shells dominated by plasma pressure, a sharp drop in luminosity is
expected. The timing of this drop is proportional to the radius of the
progenitor \citep{rabinak12}, effectively placing an upper limit to this value
\citepalias{bloom12}. At 4\,h after explosion this limit is $\lesssim
0.02$\,\Rsun \citepalias{bloom12}, but by 22\,h after explosion this rises to
$\sim0.07$\,\Rsun. Using the explosion date we estimated, the limits are similar
to those of \citetalias{nugent11}, and a factor of a few larger than those of
\citetalias{bloom12}.

\subsection{Carbon layer}
\label{sec:carbon-layer}

Our models require a carbon-dominated outermost layer, at $v > 19400$\,\kms\ 
(Fig.~\ref{fig:abundances-11fe}). The presence of this layer is supported by
the influence carbon has on the spectra: the carbon feature near 6600\,\AA\
weakens quickly, which requires carbon to be located at a high velocity.
The presence of some heavier elements,
consistent with the metal content of the progenitor, is also necessary. On the
other hand, oxygen has a negative effect on the spectra: if the outermost layer
is assumed to be composed mostly of carbon and oxygen in equal amounts, the
\OI\ 7774\,\AA\ line is too strong and blue, especially at the earliest epoch
modelled (Fig. \ref{fig:sequence-11fe-outercarbon}).  The \OI\ line in our code
matches the observations in a number of different SNe of various types,
including SNe\,Ia at early times, so we see a real difference here. The O
abundance may have been too high also in 50:50 carbon-oxygen models for other
SNe in the past, but the lack of very early data made this less obvious.  This
effect decreases with time, which again emphasizes the importance of early
data. 

This outermost layer should most directly reflect the properties of the
accreting material. There may be two scenarios in which an outer carbon-rich
layer can develop. The first is via accretion of hydrogen (in a single
degenerate system). This hydrogen can convert into carbon, but there may be no 
time for the subsequent conversion into oxygen. A second scenario is via
accretion of helium, with the helium then burning to carbon -- but this is more
unlikely as helium tends to burn explosively to heavier elements.  Thus we
qualitatively favour the hydrogen-accretion scenario, although data seem to
disfavour large companions \citep{nugent11,bloom12,Brown12} and a dense CSM
\citep{Chomiuk12,Margutti12} main sequence companions are not ruled out. 
Detecting this hydrogen, which should be located at high velocities, would be
the key to confirming this, but \HI\ lines may require non-thermal excitation
\citep{maz09,hachinger12b}, which is not very strong in the first few days after
explosion when the optical depth to the $\gamma$-rays produced by \Nifs\ decay
deep in the SN is still very large \citep{maz98}. Small amounts of hydrogen 
\citep[a few $0.01\Msun$,][]{hachinger12b}  may therefore remain undetected.
Intriguing indirect evidence for the presence of a small amount of hydrogen
comes from the very presence of HVFs, which may be explained as a consequence of
the lower ionization regime induced by the presence of some hydrogen and the
ensuing contribution to the electron density \citep{altavilla07a, tan08}.  In
SN\,2011fe HVFs are weak, but not absent \citep[see also][]{parrent12}.  The low
upper limit to the mass of H at low velocities deduced qualitatively from
nebular spectra by \citet{shappee12} appears to be a strong argument against a
H-rich companion.  This should be further investigated via detailed nebular
modelling of SN\,2011fe, which we defer to future work.

\section{Summary and Conclusions}
\label{sec:conclusions}

We have analysed a series of photospheric-epoch UV and optical spectra of the
nearby SN\,Ia SN\,2011fe. The spectra can be reproduced using a custom-made
density distribution (`$\rho$-11fe'), which has more material in the outer
layers than W7 but does not reach the densities of delayed detonation models.
Although this ad-hoc model is characterised by a lower $\KE$, this value, as
well as that of the ejected mass, which is assumed here to be the Chandrasekhar
mass, cannot be confirmed until the properties of the inner ejecta are analysed
through nebular spectroscopy. The abundance stratification in the outer layers
of SN\,2011fe is strongly reminiscent of a one-dimensional delayed detonation,
but the energy we derived for SN\,2011fe is smaller than any published DD
model. It would be interesting to verify whether a similar model can be
obtained from hydrodymamic calculations.  Also, three-dimensional delayed
detonation models of non-rotating white dwarfs, which place \Nifs\ at lower
velocities than \Feffo, may be in conflict with the late-time spectra of
SNe\,Ia \citep{seit13}. One alternative possibility is the explosion of a white
dwarf with mass slightly lower than the Chandrasekhar value, which could be
exploded by the ignition of helium accreting on its surface
\citep{shigeyama90,livne95,sim10}. This may place enough material at high
velocities.

Some \Nifs\ at relatively high velocities is required in order to reproduce the
earliest spectra of SN\,2011fe. This seems to be a common property of SNe Ia
\citep{mazzali08a}.  Stable iron and other Fe-group elements found in the
outermost layers, where burning is weak or absent, should be good tracers of
the properties of the progenitor. We derive an Fe-group content in these layers
consistent with a sub-solar metallicity of the progenitor. The relative
Fe-group abundances inferred in the incompletely-burned zones, where stable
iron and \Nifs\ are present in comparable amounts, also seems to point to a
sub-solar metallicity \citep[cf.][]{iwa99}.

\citet{piro13b} investigated the implications of different \Nifs\ abundance
distributions on the early light curve. In qualitative agreement with their
study, we find that the \Nifs\ distribution in SN\,2011fe has a high-velocity
tail. However, the abundance of \Nifs\ in the outermost 0.1\Msun\ of the ejecta
is very small. This implies that the SN had a `dark time' between the explosion
and the emergence of  significant optical light.  Spectral analysis is sensitive
to the assumed epoch because the density is a function of time. We find that a
dark time of $\sim 1.4 \pm 0.5$\,d is required. This is somewhat longer than 
the $\sim$\,0.5\,d found by \citet{piro13b}, who assume a higher abundance of
\Nifs\ in the outermost layers. This dark period has implications on the upper
limits on the radius of the progenitor ($R_\textrm{prog}$)
\citep{nugent11,bloom12}. We find $R_\textrm{prog} \lesssim 0.07$\,\Rsun.

Early UV spectra are an essential tool to explore progenitor metallicities and
understand the properties of different SNe Ia. They are much more sentitive to
metal content than the optical region. Depending on the distribution of \Nifs\
and on the presence of other Fe-group elements, the opacity of SNe with similar
luminosities can vary \citep{timmes03,mazzali06a}. This may, for example,
explain why SNe with similar luminosity show different evolution of their line
velocities \citep{hoeflich10a}. It will be important to obtain UV spectral
series for other SNe Ia in the future.

This work was made possible by the availability of a high-quality time series
of UV-optical spectra. We have inferred explosion properties, including
abundances, and crafted a density model that improves the spectra fits, in
particular in the UV. This model still has shortcomings, and it is by no means
the ultimate solution. However, it demonstrates what can be learned using
spectral modelling as a method to investigate all properties of the explosion,
not just composition. The availability of an extensive and detailed dataset
challenges theoretical models to match the observations. It would be
interesting to determine whether a model similar to ours can be obtained from
first principles. Cross-fertilisation between observations, modelling and
first-principles calculations of stellar evolution and explosion is the only
way to make real progress in our physical understanding of SNe Ia.

\section*{ACKNOWLEDGEMENTS}

This work is based on observations made with the NASA\,/\,ESA Hubble
Space Telescope, obtained at the Space Telescope Science Institute,
which is operated by the Association of Universities for Research in
Astronomy, Inc., under NASA contract NAS 5-26555. These observations
are associated with program \#12298.  Based on observations made with
the Italian Telescopio Nazionale Galileo (TNG) operated on the island
of La Palma by the Fundación Galileo Galilei of the INAF (Istituto
Nazionale di Astrofisica) at the Spanish Observatorio del Roque de los
Muchachos of the Instituto de Astrofisica de Canarias. We would like
to thank the TNG staff for their support.  PAM and SH acknowledge
support from the Italian Space Agency under programme ASI-INAF
I/009/10/0, and SH acknowlegdges further support from the Minerva
foundation (ARCHES award). MS acknowledges support from the Royal
Society.  Research by AG is supported by grants from the BSF, the EU
via an FP7/ERC grant, the ARCHES prize and the Kimmel Award for
Innovative Investigation. Collaborative work between AG and PAM is
supported by the Minerva foundation. We have made use of the NASA/IPAC
Extragalactic Database (NED,
\href{http://nedwww.ipac.caltech.edu}{http://nedwww.ipac.caltech.edu},
operated by the Jet Propulsion Laboratory, California Institute of
Technology, under contract with the National Aeronautics and Space
Administration).

\bibliographystyle{mn2e}

\begin{thebibliography}{90}
\expandafter\ifx\csname natexlab\endcsname\relax\def\natexlab#1{#1}\fi

\bibitem[{{Abbott} \& {Lucy}(1985)}]{abb85}
{Abbott} D.~C., {Lucy} L.~B., 1985, \apj, 288, 679

\bibitem[{{Altavilla} {et~al}\mbox{.}(2007){Altavilla} {et~al.}}]{altavilla07a}
{Altavilla} G., {et~al.}, 2007, \aap, 475, 585

\bibitem[{{Asplund} {et~al.}(2009)}]{asplund09} {Asplund M}., {Grevesse} N., 
{Sauval} A.~J., \& {Scott} P.\ 2009, \araa, 47, 481 

\bibitem[{{Astier}(2012)}]{astier12}
{Astier} P., 2012, Rep. Prog. Phys., 75, 116901

\bibitem[{{Balland} {et~al}\mbox{.}(2009){Balland} {et~al.}}]{balland09}
{Balland} C., {et~al.}, 2009, \aap, 507, 85

\bibitem[{{Benetti} {et~al}\mbox{.}(2005){Benetti} {et~al.}}]{ben05}
{Benetti} S., {et~al.}, 2005, \apj, 623, 1011

\bibitem[{{Bloom} {et~al}\mbox{.}(2012){Bloom} {et~al.}}]{bloom12}
{Bloom} J.~S., {et~al.}, 2012, \apjl, 744, L17

\bibitem[{{Bosma} {et~al}\mbox{.}(1981){Bosma}, {Goss}, \&
  {Allen}}]{1981A&A....93..106B}
{Bosma} A., {Goss} W.~M., {Allen} R.~J., 1981, \aap, 93, 106

\bibitem[{{Branch} \& {Venkatakrishna}(1986)}]{branch86}
{Branch} D., {Venkatakrishna} K.~L., 1986, \apjl, 306, L21

\bibitem[{{Brown} {et~al}\mbox{.}(2012){Brown} {et~al.}}]{Brown12} 
{Brown} P.~J., et al.\ 2012, \apj, 753, 22 

\bibitem[{{Bufano} {et~al}\mbox{.}(2009){Bufano} {et~al.}}]{bufano09}
{Bufano} F., {et~al.}, 2009, \apj, 700, 1456

\bibitem[{{Cappellaro} {et~al}\mbox{.}(1995){Cappellaro}, {Turatto}, \&
  {Fernley}}]{cappellaro95}
{Cappellaro} E., {Turatto} M., {Fernley} J., 1995, ESA-SP 1189

\bibitem[{{Cardelli} {et~al}\mbox{.}(1989){Cardelli}, {Clayton}, \&
  {Mathis}}]{car89}
{Cardelli} J.~A., {Clayton} G.~C., {Mathis} J.~S., 1989, \apj, 345, 245

\bibitem[{{Chomiuk} {et~al.}(2012){Chomiuk} {et~al.}}]{Chomiuk12} 
Chomiuk L., et al.\ 2012, \apj, 750, 164 

 
\bibitem[{{Conley} {et~al}\mbox{.}(2008){Conley} {et~al.}}]{conley08}
{Conley} A., {et~al.}, 2008, \apj, 681, 482

\bibitem[{{Conley} {et~al}\mbox{.}(2011){Conley}, {Guy}, {Sullivan},
  {Regnault}, {Astier}, {Balland}, {Basa}, {Carlberg}, {et~al.}}]{conley11}
{Conley} A. {et~al.}, 2011, \apjs, 192, 1

\bibitem[{{de Vaucouleurs} {et~al}\mbox{.}(1991){de Vaucouleurs}, {de
  Vaucouleurs}, {Corwin}, {Buta}, {Paturel}, \& 
  {Fouqu{\'e}}}]{1991rc3..book.....D}
{de Vaucouleurs} G., {de Vaucouleurs} A., {Corwin} Jr. H.~G., {Buta} R.~J.,
  {Paturel} G., {Fouqu{\'e}} P., 1991, {Third Reference Catalogue of Bright
  Galaxies. Vol. III.} Spinger, New York, (USA)

\bibitem[{{Ellis} {et~al}\mbox{.}(2008){Ellis} {et~al.}}]{ellis08}
{Ellis} R.~S., {et~al.}, 2008, \apj, 674, 51

\bibitem[{{Foley} \& {Kirshner}(2013)}]{foley13a}
{Foley} R.~J., {Kirshner} R.~P., 2013, \apjl, 769, L1 

\bibitem[{{Foley} {et~al}\mbox{.}(2008{\natexlab{a}}){Foley}, {Filippenko},
  {Aguilera}, {Becker}, {Blondin}, {Challis}, {Clocchiat ti}, {Covarrubias},
  {Davis}, {Garnavich}, {Jha}, {Kirshner}, {Krisciunas}, {Leibundgut}, {Li},
  {Matheson}, {Miceli}, {Miknaitis}, {Pignata}, {Rest}, {Riess}, {Schmidt},
  {Smith}, {Sollerman}, {Spyromilio}, {Stubbs}, {Suntzeff}, {Tonry},
  {Wood-Vasey}, \& {Zenteno}}]{foley08a}
{Foley} R.~J. {et~al.}, 2008{\natexlab{a}}, \apj, 684, 68

\bibitem[{{Foley} {et~al}\mbox{.}(2008{\natexlab{b}}){Foley}, {Filippenko}, \&
  {Jha}}]{foley08b}
{Foley} R.~J., {Filippenko} A.~V., {Jha} S.~W., 2008{\natexlab{b}}, \apj, 686,
  117

\bibitem[{{Foley} {et~al}\mbox{.}(2012{\natexlab{a}}){Foley}
  {et~al.}}]{foley12sdss}
{Foley} R.~J., {et~al.}, 2012{\natexlab{a}}, \aj, 143, 113

\bibitem[{{Foley} {et~al}\mbox{.}(2012{\natexlab{b}}){Foley}
  {et~al.}}]{foley12sn2011iv}
{Foley} R.~J., {et~al.}, 2012{\natexlab{b}}, \apjl, 753, L5

\bibitem[{{Foley} {et~al}\mbox{.}(2012{\natexlab{c}}){Foley}
  {et~al.}}]{foley12sn2009ig}
{Foley} R.~J., {et~al.}, 2012{\natexlab{c}}, \apj, 744, 38

\bibitem[{{Guy} {et~al}\mbox{.}(2007){Guy}, {Astier}, {Baumont}, {Hardin},
  {Pain}, {Regnault}, {Basa}, {Carlberg}, {Conley}, {Fabbro}, {Fouchez},
  {Hook}, {Howell}, {Perrett}, {Pritchet}, {Rich}, {Sullivan}, {Antilogus},
  {Aubourg}, {Bazin}, {Bronder}, {Filiol}, {Palanque-Delabrouille}, {Ripoche},
  \& {Ruhlmann-Kleider}}]{guy07}
{Guy} J. {et~al.}, 2007, \aap, 466, 11

\bibitem[{{Hachinger} {et~al}\mbox{.}(2008){Hachinger}, {Mazzali}, {Tanaka},
  {Hillebrandt}, \& {Benetti}}]{hac08}
{Hachinger} S., {Mazzali} P.~A., {Tanaka} M., {Hillebrandt} W., {Benetti} S.,
  2008, \mnras, 389, 1087

\bibitem[{{Hachinger} {et~al}\mbox{.}(2009){Hachinger}, {Mazzali},
  {Taubenberger}, {Pakmor}, \& {Hillebrandt}}]{hachinger09}
{Hachinger} S., {Mazzali} P.~A., {Taubenberger} S., {Pakmor} R., {Hillebrandt}
  W., 2009, \mnras, 399, 1238
  
\bibitem[{{Hachinger} {et~al}\mbox{.}(2012){Hachinger}, {Mazzali},
  {Taubenberger}, {Fink}, {Pakmor}, {Hillebrandt}, \&
  {Seitenzahl}}]{hachinger12b}
{Hachinger} S., {Mazzali} P.~A., {Taubenberger} S., {Fink} M., {Pakmor} R.,
  {Hillebrandt} W., {Seitenzahl} I.~R., 2012, \mnras, 427, 2057

\bibitem[{{Hachinger} {et~al}\mbox{.}(2013){Hachinger} {et~al.}}]{hachinger13}
{Hachinger} S., {et~al.}, 2013, \mnras, 429, 2228

\bibitem[{{Hayden} {et~al}\mbox{.}(2010){Hayden} {et~al.}}]{hayden10a}
{Hayden} B.~T., {et~al.}, 2010, \apj, 712, 350

\bibitem[{{Hoeflich} {et~al}\mbox{.}(1998){Hoeflich}, {Wheeler}, \&
  {Thielemann}}]{hoe98}
{Hoeflich} P., {Wheeler} J.~C., {Thielemann} F.~K., 1998, \apj, 495, 617

\bibitem[{{H{\"o}flich} {et~al}\mbox{.}(2004){H{\"o}flich}, {Gerardy},
  {Nomoto}, {Motohara}, {Fesen}, {Maeda}, {Ohkubo}, {Tominaga}}]{hoeflich04}
{H{\"o}flich} P., {Gerardy} C.~L., {Nomoto} K., {Motohara} K., {Fesen} R.~A.,
  {Maeda} K., {Ohkubo} T., {Tominaga} N., 2004, \apj, 617, 1258

\bibitem[{{H{\"o}flich} {et~al}\mbox{.}(2010){H{\"o}flich}, {Krisciunas},
  {Khokhlov}, {Baron}, {Folatelli}, {Hamuy}, {Phillips}, {Suntzeff},
  {et~al.}}]{hoeflich10a}
{H{\"o}flich} P. {et~al.}, 2010, \apj, 710, 444

\bibitem[{{Howell}(2011)}]{howell11}
{Howell} D.~A., 2011, Nature Communications, 2, 350

\bibitem[{{Hsiao} {et~al}\mbox{.}(2013){Hsiao}, {Marion}, {Phillips}, {Burns},
  {Winge}, {Morrell}, {Contreras}, {Freedman}, {Kromer}, {Gall}, {Gerardy},
  {H{\"o}flich}, {Im}, {Jeon}, {Kirshner}, {Nugent}, {Persson}, {Pignata},
  {Roth}, {Stanishev}, {Stritzinger}, \& {Suntzeff}}]{hsiao13}
{Hsiao} E.~Y. {et~al.}, 2013, \apj, 766, 72

\bibitem[{{Iwamoto} {et~al}\mbox{.}(1999){Iwamoto}, {Brachwitz}, {Nomoto},
  {Kishimoto}, {Umeda}, {Hix}, \& {Thielemann}}]{iwa99}
{Iwamoto} K., {Brachwitz} F., {Nomoto} K., {Kishimoto} N., {Umeda} H., {Hix}
  W.~R., {Thielemann} F., 1999, \apjs, 125, 439

\bibitem[{{Jeffery} {et~al}\mbox{.}(1992){Jeffery}, {Leibundgut}, {Kirshner},
  {Benetti}, {Branch}, \& {Sonneborn}}]{jeffrey92}
{Jeffery} D.~J., {Leibundgut} B., {Kirshner} R.~P., {Benetti} S., {Branch} D.,
  {Sonneborn} G., 1992, \apj, 397, 304

\bibitem[{{Kasen}(2010)}]{kas10}
{Kasen} D., 2010, \apj, 708, 1025

\bibitem[{{Kessler} {et~al}\mbox{.}(2009){Kessler} {et~al.}}]{kes09}
{Kessler} R., {et~al.}, 2009, \apjs, 185, 32

\bibitem[{{Khokhlov}(1991)}]{kho91a}
{Khokhlov} A.~M., 1991, \aap, 245, 114

\bibitem[{{Kirshner} {et~al}\mbox{.}(1993){Kirshner}, {Jeffery}, {Leibundgut},
  {Challis}, {Sonneborn}, {Phillips}, {Suntzeff}, {Smith},
  {et~al.}}]{kirshner93}
{Kirshner} R.~P. {et~al.}, 1993, \apj, 415, 589

\bibitem[{{Law} {et~al}\mbox{.}(2009){Law} {et~al.}}]{law09}
{Law} N.~M., {et~al.}, 2009, \pasp, 121, 1395

\bibitem[{{Leibundgut} {et~al}\mbox{.}(1991){Leibundgut}, {Kirshner},
  {Filippenko}, {Shields}, {Foltz}, {Phillips}, \& {Sonneborn}}]{leibundgut91}
{Leibundgut} B., {Kirshner} R.~P., {Filippenko} A.~V., {Shields} J.~C., {Foltz}
  C.~B., {Phillips} M.~M., {Sonneborn} G., 1991, \apjl, 371, L23

\bibitem[{{Lentz} {et~al}\mbox{.}(2000){Lentz}, {Baron}, {Branch},
  {Hauschildt}, \& {Nugent}}]{lentz00a}
{Lentz} E.~J., {Baron} E., {Branch} D., {Hauschildt} P.~H., {Nugent} P.~E.,
  2000, \apj, 530, 966

\bibitem[{{Livne} \& {Arnett}(1995)}]{livne95} Livne, E., \& Arnett, D.\ 1995,
\apj, 452, 62 

\bibitem[{{Lucy}(1999)}]{luc99}
{Lucy} L.~B., 1999, \aap, 345, 211

\bibitem[{{Maguire} {et~al}\mbox{.}(2012){Maguire}, {Sullivan}, {Ellis},
  {Nugent}, {Howell}, {Gal-Yam}, {Cooke}, {Mazzali}, {Pan}, {Dilday}, {Thomas},
  {Arcavi}, {Ben-Ami}, {Bersier}, {Bianco}, {Fulton}, {Hook}, {Horesh},
  {Hsiao}, {James}, {Podsiadlowski}, {Walker}, {Yaron}, {Kasliwal}, {Laher},
  {Law}, {Ofek}, {Poznanski}, \& {Surace}}]{maguire12}
{Maguire} K. {et~al.}, 2012, \mnras, 426, 2359

\bibitem[{{Margutti} {et~al}\mbox{.}(2012){Margutti} {et~al.}}]{Margutti12} 
Margutti R., et al.\ 2012, \apj, 751, 134 

\bibitem[{{Mazzali}(2000)}]{maz00}{Mazzali} P.~A., 2000, \aap, 363, 705

\bibitem[{{Mazzali}(2001)}]{mazzali2001}{Mazzali} P.~A., 2001, \mnras, 321, 341 

\bibitem[{{Mazzali} \& {Lucy}(1993)}]{mazzali93b}
{Mazzali} P.~A., {Lucy} L.~B., 1993, \aap, 279, 447

\bibitem[{{Mazzali} \& {Lucy}(1998)}]{maz98}
{Mazzali} P.~A., {Lucy} L.~B., 1998, \mnras, 295, 428

\bibitem[{{Mazzali} \& {Schmidt}(2005)}]{mazz_sch05}
{Mazzali} P.~A., {Schmidt} B.~P., 2005, in IAU Symposium, Vol. 201, New
  Cosmological Data and the Values of the Fundamental Parameters, {Lasenby}
  A.~N., {Wilkinson} A., eds., p. 241

\bibitem[{{Mazzali} \& {Podsiadlowski}(2006)}]{mazzali06a}
{Mazzali} P.~A., {Podsiadlowski} P., 2006, \mnras, 369, L19

\bibitem[{{Mazzali} \& {Hachinger}(2012)}]{mazzali12}
{Mazzali} P.~A., {Hachinger} S., 2012, \mnras, 424, 2926

\bibitem[{{Mazzali} {et~al}\mbox{.}(1993){Mazzali}, {Lucy}, {Danziger},
  {Gouiffes}, {Cappellaro}, \& {Turatto}}]{mazzali93a}
{Mazzali} P.~A., {Lucy} L.~B., {Danziger} I.~J., {Gouiffes} C., {Cappellaro}
  E., {Turatto} M., 1993, \aap, 269, 423

\bibitem[{{Mazzali} {et~al.}(1998)}]{mazzali98} {Mazzali} P.~A., 
{Cappellaro} E., {Danziger} I.~J., {Turatto} M., 
\& {Benetti} S.\ 1998, \apjl, 499, L49 

\bibitem[{{Mazzali} {et~al}\mbox{.}(2001){Mazzali}, {Nomoto}, {Cappellaro},
  {Nakamura}, {Umeda}, \& {Iwamoto}}]{maz01lc}
{Mazzali} P.~A., {Nomoto} K., {Cappellaro} E., {Nakamura} T., {Umeda} H.,
  {Iwamoto} K., 2001, \apj, 547, 988

\bibitem[{{Mazzali} {et~al}\mbox{.}(2005){Mazzali} {et~al.}}]{mazzali05b}
{Mazzali} P.~A., {et~al.}, 2005, \apjl, 623, L37

\bibitem[{{Mazzali} {et~al}\mbox{.}(2007){Mazzali}, {R{\"o}pke}, {Benetti}, \&
  {Hillebrandt}}]{maz07}
{Mazzali} P.~A., {R{\"o}pke} F.~K., {Benetti} S., {Hillebrandt} W., 2007, \sci,
  315, 825

\bibitem[{{Mazzali} {et~al}\mbox{.}(2009){Mazzali}, {Deng}, {Hamuy}, \&
  {Nomoto}}]{maz09}
{Mazzali} P.~A., {Deng} J., {Hamuy} M., {Nomoto} K., 2009, \apj,
  703, 1624
 
\bibitem[{{Mazzali} {et~al}\mbox{.}(2008){Mazzali}, {Sauer}, {Pastorello},
  {Benetti}, \& {Hillebrandt}}]{mazzali08a}
{Mazzali} P.~A., {Sauer} D.~N., {Pastorello} A., {Benetti} S., {Hillebrandt}
  W., 2008, \mnras, 386, 1897

\bibitem[{{Mazzali} {et~al}\mbox{.}(2011){Mazzali}, {Maurer}, {Stritzinger},
  {Taubenberger}, {Benetti}, \& {Hachinger}}]{mazzali11}
{Mazzali} P.~A., {Maurer} I., {Stritzinger} M., {Taubenberger} S., {Benetti}
  S., {Hachinger} S., 2011, \mnras, 416, 881

\bibitem[{{Munari} {et~al}\mbox{.}(2013){Munari}, {Henden}, {Belligoli},
  {Castellani}, {Cherini}, {Righetti}, \& {Vagnozzi}}]{munari13}
{Munari} U., {Henden} A., {Belligoli} R., {Castellani} F., {Cherini} G.,
  {Righetti} G.~L., {Vagnozzi} A., 2013, \na, 20, 30

\bibitem[{{Nomoto} {et~al}\mbox{.}(1984){Nomoto}, {Thielemann}, \&
  {Yokoi}}]{nom84w7}
{Nomoto} K., {Thielemann} F., {Yokoi} K., 1984, \apj, 286, 644

\bibitem[{{Nugent} {et~al}\mbox{.}(2011){Nugent}, {Sullivan}, {Cenko},
  {Thomas}, {Kasen}, {Howell}, {Bersier}, {Bloom}, {et~al.}}]{nugent11}
{Nugent} P.~E. {et~al.}, 2011, \nat, 480, 344

\bibitem[{{Parrent} {et~al}\mbox{.}(2012){Parrent} {et~al.}}]{parrent12}
{Parrent} J.~T., {et~al.}, 2012, \apjl, 752, L26

\bibitem[{{Patat} {et~al}\mbox{.}(2013){Patat} {et~al.}}]{patat13}
{Patat} F., {et~al.}, 2013, \aap, 549, A62

\bibitem[{{Pereira} {et~al}\mbox{.}(2013){Pereira} {et~al.}}]{pereira13}
{Pereira} R., {et~al.}, 2013, \aap, 554, A27 

\bibitem[{{Perlmutter} {et~al}\mbox{.}(1999){Perlmutter} {et~al.}}]{per99}
{Perlmutter} S., {et~al.}, 1999, \apj, 517, 565

\bibitem[{{Piro} \& {Nakar}(2013{\natexlab{a}})}]{piro13a}
{Piro} A.~L., {Nakar} E., 2013{\natexlab{a}}, \apj, 769, 67 

\bibitem[{{Piro} \& {Nakar}(2013{\natexlab{b}})}]{piro13b}
{Piro} A.~L., {Nakar} E., 2013{\natexlab{b}}, \apj, submitted (arXiv:1211.6438)

\bibitem[{{Rabinak} \& {Waxman}(2011)}]{rabinak11}
{Rabinak} I., {Waxman} E., 2011, \apj, 728, 63

\bibitem[{{Rabinak} {et~al}\mbox{.}(2012){Rabinak}, {Livne}, \&
  {Waxman}}]{rabinak12}
{Rabinak} I., {Livne} E., {Waxman} E., 2012, \apj, 757, 35

\bibitem[{{Rau} {et~al}\mbox{.}(2009){Rau} {et~al.}}]{rau09}
{Rau} A., {et~al.}, 2009, \pasp, 121, 1334

\bibitem[{{Richmond} \& {Smith}(2012)}]{richmond12}
{Richmond} M.~W., {Smith} H.~A., 2012, Journal of the American Association of
  Variable Star Observers (JAAVSO), 40, 872

\bibitem[{{Riess} {et~al}\mbox{.}(1998){Riess} {et~al.}}]{rie98}
{Riess} A.~G., {et~al.}, 1998, \aj, 116, 1009

\bibitem[{{Riess} {et~al}\mbox{.}(2007){Riess} {et~al.}}]{rie07}
{Riess} A.~G., {et~al.}, 2007, \apj, 659, 98

\bibitem[{{R{\"o}pke} {et~al}\mbox{.}(2007){R{\"o}pke}, {Hillebrandt},
  {Schmidt}, {Niemeyer}, {Blinnikov}, \& {Mazzali}}]{roe07b}
{R{\"o}pke} F.~K., {Hillebrandt} W., {Schmidt} W., {Niemeyer} J.~C.,
  {Blinnikov} S.~I., {Mazzali} P.~A., 2007, \apj, 668, 1132

\bibitem[{{R{\"o}pke} {et~al}\mbox{.}(2012){R{\"o}pke}, {Kromer}, {Seitenzahl},
  {Pakmor}, {Sim}, {Taubenberger}, {Ciaraldi-Schoolmann}, {Hillebrandt},
  {et~al.}}]{roepke12}
{R{\"o}pke} F.~K. {et~al.}, 2012, \apjl, 750, L19

\bibitem[{{Sauer} {et~al}\mbox{.}(2008{\natexlab{a}}){Sauer}, {Mazzali},
  {Blondin}, {Stehle}, {Benetti}, {Challis}, {Filippenko}, {Kirshner}, {Li}, \&
  {Matheson}}]{sauer08}
{Sauer} D.~N. {et~al.}, 2008{\natexlab{a}}, \mnras, 391, 1605

\bibitem[{{Sauer} {et~al}\mbox{.}(2008{\natexlab{b}}){Sauer} {et~al.}}]{sau08}
{Sauer} D.~N., {et~al.}, 2008{\natexlab{b}}, \mnras, 391, 1605

\bibitem[{{Schlegel} {et~al}\mbox{.}(1998){Schlegel}, {Finkbeiner}, \&
  {Davis}}]{sch98}
{Schlegel} D.~J., {Finkbeiner} D.~P., {Davis} M., 1998, \apj, 500, 525

\bibitem[{{Seitenzahl} {et~al.}(2013)}]{seit13} Seitenzahl, I.~R., 
Ciaraldi-Schoolmann, F., R{\"o}pke, F.~K., et al.\ 2013, \mnras, 429, 1156
 
\bibitem[{{Shappee} \& {Stanek}(2011)}]{shappee11}
{Shappee} B.~J., {Stanek} K.~Z., 2011, \apj, 733, 124

\bibitem[{{Shappee} {et~al.}(2013)}]{shappee12} Shappee, B.~J., Stanek, 
K.~Z., Pogge, R.~W., \& Garnavich, P.~M.\ 2013, \apjl, 762, L5 

\bibitem[{{Shigeyama} {et al.}(1990)}]{shigeyama90} Shigeyama, T., 
Nomoto, K., Tsujimoto, T., \& Hashimoto, M.-A.\ 1990, \apjl, 361, L23 

\bibitem[{{Sim} {et al.}(2010)}]{sim10} Sim, S.~A., R{\"o}pke, 
F.~K., Hillebrandt, W., et al.\ 2010, \apjl, 714, L52 

\bibitem[{{Steele} {et~al}\mbox{.}(2004){Steele} {et~al.}}]{ste04}
{Steele} I.~A., {et~al.}, 2004, in SPIE Conference Series, {Oschmann} Jr.
  J.~M., ed., Vol. 5489, SPIE, Bellingham WA, p. 679

\bibitem[{{Stehle} {et~al}\mbox{.}(2005){Stehle}, {Mazzali}, {Benetti}, \&
  {Hillebrandt}}]{ste05}
{Stehle} M., {Mazzali} P.~A., {Benetti} S., {Hillebrandt} W., 2005, \mnras,
  360, 1231

\bibitem[{{Stoll} {et~al}\mbox{.}(2011){Stoll}, {Shappee}, \&
  {Stanek}}]{stoll11}
{Stoll} R., {Shappee} B., {Stanek} K.~Z., 2011, ATEL 3588

\bibitem[{{Stritzinger} {et~al}\mbox{.}(2006){Stritzinger}, {Mazzali},
  {Sollerman}, \& {Benetti}}]{stritzinger06b}
{Stritzinger} M., {Mazzali} P.~A., {Sollerman} J., {Benetti} S., 2006, \aap,
  460, 793

\bibitem[{{Sullivan} {et~al}\mbox{.}(2011){Sullivan} {et~al.}}]{sul11}
{Sullivan} M., {et~al.}, 2011, \apj, 737, 102

\bibitem[{{Suzuki} {et~al}\mbox{.}(2012){Suzuki} {et~al.}}]{suz12}
{Suzuki} N., {et~al.}, 2012, \apj, 746, 85

\bibitem[{{Tanaka} {et~al}\mbox{.}(2006){Tanaka} {et~al.}}]{tanaka06}
{Tanaka} M., {Mazzali} P.~A., {Maeda} K., {Nomoto} K., 2006, \apj, 645, 470

\bibitem[{{Tanaka} {et~al}\mbox{.}(2008){Tanaka} {et~al.}}]{tan08}
{Tanaka} M., {et~al.}, 2008, \apj, 677, 448

\bibitem[{{Tanaka} {et~al}\mbox{.}(2011){Tanaka}, {Mazzali}, {Stanishev},
  {Maurer}, {Kerzendorf}, \& {Nomoto}}]{tanaka11}
{Tanaka} M., {Mazzali} P.~A., {Stanishev} V., {Maurer} I., {Kerzendorf} W.~E.,
  {Nomoto} K., 2011, \mnras, 410, 1725

\bibitem[{{Timmes} {et~al}\mbox{.}(2003){Timmes}, {Brown}, \&
  {Truran}}]{timmes03}
{Timmes} F.~X., {Brown} E.~F., {Truran} J.~W., 2003, \apjl, 590, L83

\bibitem[{{Vink{\'o}} {et~al}\mbox{.}(2012){Vink{\'o}} {et~al.}}]{vinko12}
{Vink{\'o}} J., {et~al.}, 2012, \aap, 546, A12

\bibitem[{{Walker} {et~al}\mbox{.}(2012){Walker} {et~al.}}]{wal12}
{Walker} E., {et~al.}, 2012, MNRAS, 427, 103

\bibitem[{{Wang} {et~al}\mbox{.}(2012){Wang} {et~al.}}]{wang12}
{Wang} X., {et~al.}, 2012, \apj, 749, 126

\bibitem[{{Yaron} \& {Gal-Yam}(2012)}]{yaron12}
{Yaron} O., {Gal-Yam} A., 2012, \pasp, 124, 668

\end{thebibliography}

\end{document}